\documentclass[11pt]{article}
\usepackage[a4paper]{geometry}
\usepackage{abstract}
\usepackage[utf8]{inputenc}
\usepackage{array}
\usepackage{float}
\usepackage{booktabs}
\usepackage{mathtools}
\usepackage{multirow}
\usepackage{amsmath}
\usepackage{amsthm}
\usepackage{amssymb}
\usepackage{graphicx}
\usepackage[authoryear]{natbib}
\usepackage[unicode=true,
 bookmarks=false,
 breaklinks=false,pdfborder={0 0 1},backref=section,colorlinks=false]
 {hyperref}

\makeatletter

\newtheorem{remark}{Remark}

\usepackage{bm}
\usepackage{algorithm}
\usepackage{algorithmic}
\usepackage{authblk}
\usepackage[inline]{enumitem}
\usepackage{subcaption}
\usepackage{color}
\usepackage{cleveref}
\newtheorem{theorem}{Theorem}\newtheorem{proposition}{Proposition}\newtheorem{assumption}{Assumption}
\newtheorem{corollary}{Corollary}

\title{Breaking the Dimensional Barrier for Constrained Dynamic Portfolio Choice}

\author[1]{Jaegi Jeon}
\author[2]{Jeonggyu Huh\thanks{Corresponding Author: jghuh@skku.edu}}
\author[3]{Hyeng Keun Koo}
\author[4]{Byung Hwa Lim}
\affil[1]{\small Graduate School of Data Science, Chonnam National University, Gwangju, Republic of Korea}
\affil[2]{\small Department of Mathematics, Sungkyunkwan University, Suwon, Republic of Korea}
\affil[3]{\small Department of Financial Engineering, Ajou University, Suwon, Republic of Korea}
\affil[4]{\small Department of Fintech, SKK Business School, Sungkyunkwan University, Seoul, Republic of Korea}

\makeatother

\begin{document}

\maketitle

\begin{abstract}
We propose a scalable, policy‑centric framework for continuous‑time multi‑asset portfolio–consumption optimization under inequality constraints. Our method integrates neural policies with Pontryagin’s Maximum Principle (PMP) and enforces feasibility by maximizing a log‑barrier–regularized Hamiltonian at each time–state pair, thereby satisfying KKT conditions without value‑function grids. Theoretically, we show that the barrier‑regularized Hamiltonian yields O($\epsilon$) policy error and a linear Hamiltonian gap (quadratic when the KKT solution is interior), and we extend the BPTT–PMP correspondence to constrained settings with stable costate convergence. Empirically, PG‑DPO and its projected variant (P‑PGDPO) recover KKT‑optimal policies in canonical short‑sale and consumption‑cap problems while maintaining strict feasibility across dimensions; unlike PDE/BSDE solvers, runtime scales linearly with the number of assets and remains practical at n=100. These results provide a rigorous and scalable foundation for high‑dimensional constrained continuous‑time portfolio optimization.
\end{abstract}

{\small
{\em Keywords:} Constrained continuous-time portfolio optimization; Pontryagin's Maximum Principle (PMP); Barrier-regularized Hamiltonian; Karush--Kuhn--Tucker (KKT) conditions; High-dimensional scalability
}

\section{Introduction}\label{sec:intro}

Dynamic portfolio choice—how an investor allocates wealth across multiple assets and consumes over time under uncertainty—has been a cornerstone of modern financial economics since the seminal works of \citet{samuelson1975lifetime} and \citet{merton1969lifetime,merton1971optimum}. Under idealized assumptions such as complete markets, frictionless trading, and unlimited shorting or borrowing, this “Merton problem” admits elegant closed-form or semi-analytic solutions, even with stochastic investment opportunities \citep[e.g.,][]{kim1996dynamic,liu2007portfolio}. In reality, however, investors face a range of frictions and regulatory constraints—short-sale bans, leverage limits, pointwise consumption bounds, and ratcheting or drawdown rules—that invalidate the frictionless formulation. These features transform the Bellman equation into a variational inequality or free-boundary problem \citep[e.g.,][]{karatzas1987optimal,cvitanic1993hedging,karatzas1998methods,XuShreve1992a,XuShreve1992b,fleming2006controlled,dybvig1995dusenberry,constantinides1990habit,sundaresan1989intertemporally}, rendering analytical solutions intractable. Classical dynamic programming (DP) methods also suffer from the curse of dimensionality, limiting empirical and numerical analysis to low-dimensional settings even without constraints \citep{campbell1999consumption,campbell2001should,campbell2003multivariate,balduzzi1999transaction,lynch2000predictability,lynch2001portfolio,brandt2005simulation,buraschi2010correlation,garlappi2010solving,jurek2011optimal}. Related interior-point and variational-inequality formulations tailored to constrained portfolio problems retain the same tractability barriers in higher dimensions due to active-set identification and free-boundary effects; see, for example, \citet{yang2019constrained} in addition to the classical treatments above.

Recent advances in deep learning have renewed interest in continuous-time stochastic control by enabling neural approximations to high-dimensional partial differential equations (PDEs) and backward stochastic differential equations (BSDEs). Yet, most existing approaches—such as deep BSDE networks \citep{han2018solving, e2017deep} and physics-informed neural networks (PINNs; \citealp{raissi2019physics, dai2023learning})—remain value-based: they approximate the value function or its gradients rather than the control itself. These methods become numerically unstable when inequality constraints are present. For example, reflected BSDEs require learning the free-boundary process enforcing feasibility, which is notoriously difficult beyond low dimensions \citep[]{el1997reflected,krishnapriyan2021characterizing}. As a result, realistic constrained portfolio problems remain unsolved at scale, with most PDE- or BSDE-based methods failing beyond five to ten assets. Closer to our setting, \citet{davey2022deep} develop a 2BSDE-style solver that scales to portfolios with roughly fifty assets; however, their analysis focuses on investment policies alone and does not incorporate consumption or pointwise inequality constraints.

To overcome these limitations, we propose a fundamentally different, \emph{policy-centric} approach that learns directly in the space of controls. Following \citet{huh2025breaking}, our \emph{Pontryagin-Guided Direct Policy Optimization} (PG-DPO) algorithm integrates Pontryagin’s Maximum Principle (PMP) into a differentiable training pipeline: neural consumption and investment policies are updated via backpropagation-through-time (BPTT), with adjoint (costate) processes steering gradient updates toward Hamiltonian stationarity. This approach eliminates the need for high-dimensional value-function grids and free-boundary estimation. Building on this foundation, we introduce \emph{Projected PG-DPO} (P-PGDPO), a constrained extension that enforces portfolio and consumption feasibility through a barrier-regularized Hamiltonian satisfying the Karush–Kuhn–Tucker (KKT) first-order conditions at every time–state pair. The term “projection” refers to satisfaction of these first-order conditions, rather than a metric (e.g., Euclidean) projection.

Theoretically, our framework establishes two main results. First, we prove a \emph{barrier–KKT correspondence theorem}, showing that the barrier-regularized Hamiltonian achieves $O(\epsilon)$ policy accuracy and an $O(\epsilon^2)$ instantaneous Hamiltonian gap, thereby linking the interior-point barrier path to the exact KKT manifold \citep{fleming2006controlled,pham2009continuous}. Second, we extend the BPTT–PMP correspondence to constrained settings, demonstrating that discrete autodifferentiated costates converge to the continuous-time adjoint processes even under inequality constraints. Together, these results form the first rigorous bridge between continuous-time PMP theory and end-to-end neural policy optimization under realistic constraints.

Empirically, we validate the framework in two canonical constrained settings: a short-sale ban on risky assets and a wealth-proportional consumption cap. Across all experiments, the proposed methods accurately recover the analytic KKT-optimal policies while maintaining strict feasibility. The barrier-projected variant (P-PGDPO) reduces policy errors by up to two orders of magnitude relative to activation-only training and remains numerically stable in large-scale environments with 100 or more assets. This scalability represents a fundamental advance over value-based solvers, which become computationally infeasible even at moderate dimensions.

In summary, our contributions are twofold. \emph{Methodologically}, we introduce a PMP-based policy-centric optimization framework that unifies barrier regularization, KKT feasibility, and BPTT adjoint consistency within a single learning paradigm. \emph{Practically}, we demonstrate that this approach delivers, to our knowledge, the first accurate, feasible numerical solutions to high-dimensional constrained continuous-time portfolio problems—precisely in cases where classical DP, PDE, or BSDE methods fail.

The remainder of the paper is organized as follows. Section~\ref{sec:dp_multiasset} revisits the DP and HJB formulation of the multi-asset constrained Merton problem and outlines their limitations. Section~\ref{sec:PMP_multiasset} introduces the PMP formulation, develops the barrier-regularized Hamiltonian framework, and establishes its correspondence with the KKT system. Section~\ref{sec:BPTT_multiasset} presents the proposed learning algorithms, PG-DPO and P-PGDPO, and establishes a BPTT–PMP correspondence with a policy-gap decomposition. Section~\ref{sec:num_test} reports numerical experiments on constrained portfolio and consumption problems. Section~\ref{sec:conclusion} concludes.

\section{Multi-Asset Portfolio Problems under Constraints}
\label{sec:dp_multiasset}

\subsection{Model Setup}
\label{sec:dp_unconstrained}

Dynamic portfolio choice describes how an investor allocates wealth across risky and risk-free assets and decides how much to consume over time. We consider a continuous-time economy in which the investment opportunities are constant, following the classic setup of Merton's model \citep{merton1969lifetime,merton1971optimum} but allowing for realistic trading and consumption constraints.\footnote{See \citet{huh2025breaking} for a complementary neural-network approach to dynamic portfolio choice with stochastically varying investment opportunities, but in an unconstrained setting without trading frictions or consumption limits.
  }
In particular, we consider an investor who allocates wealth among one risk-free asset and multiple risky assets. Let $r$ denote the constant risk-free rate of riskless asset $S_0$. There are $n$ risky assets, $\mathbf{S}\in \mathbb{R}^n$, with a constant drift vector $\boldsymbol{\mu}$ and a covariance matrix $\boldsymbol{\Sigma}=\boldsymbol{v}\boldsymbol{v}^\top$, where $\boldsymbol{v}$ is an $n\times n$ volatility factorization. In vector form, we define
$
\widetilde{\boldsymbol{\mu}}
=
\bigl(r,\;\boldsymbol{\mu}^\top\bigr)^\top,$ $
\widetilde{\boldsymbol{v}}
=
\begin{bmatrix}
\mathbf{0}_{1\times n}\\
\boldsymbol{v}
\end{bmatrix},
$
so that $\boldsymbol{\Sigma} = \boldsymbol{v}\boldsymbol{v}^\top$, with the first row of $\widetilde{\boldsymbol{v}}$ equal to zero, reflecting the zero volatility of the risk-free asset. We suppose that the covariance $\boldsymbol{\Sigma}$ is uniformly positive definite so that there exists $\underline{\sigma}>0$ such that $\boldsymbol{\xi}^\top\boldsymbol{\Sigma}\boldsymbol{\xi}\geq\underline{\sigma}\|\boldsymbol{\xi}\|^2$ almost surely.

Let $X_t$ be the total wealth of the investor at time $t$. The investor chooses a portfolio vector $\boldsymbol{\pi}_t=(\pi_{0,t},\pi_{1,t},\dots,\pi_{n,t})^\top$, representing the proportion of wealth invested in each asset, satisfying $\sum_{i=0}^n \pi_{i,t}=1$, and a consumption rate $C_t\geq0$. The wealth process evolves according to the standard continuous-time self-financing condition:  
\[
dX_t = \left(X_t\,\boldsymbol{\pi}_t^\top \widetilde{\boldsymbol{\mu}} - C_t\right) dt 
+ X_t\,\boldsymbol{\pi}_t^\top\,\widetilde{\boldsymbol{v}} \,d\mathbf{W}_t,
\qquad 
X_0=x_0>0,
\]
where $\mathbf{W}_t=(W_{1,t}, ..., W_{n,t})^\top$ is an $\mathbb{R}^n$-valued standard Brownian motion adapted to the filtration $\mathcal{F}=\{\mathcal{F}_t\}$ under the filtered probability space $(\Omega, \mathcal{F}, \mathbb{P})$. 

The investor's aim is to maximize the expected discounted utility from consumption over time and terminal wealth at the planning horizon $T$:
\[
J(\boldsymbol{\pi},C)=\mathbb{E}\left[\int_0^T e^{-\rho t}U(C_t)\,dt 
+ K e^{-\rho T}U(X_T)\right],
\]
where $\rho>0$ is the subjective discount rate and $K\ge0$ determines the relative weight placed on terminal wealth. This objective balances the desire for current consumption against long-run capital accumulation. We assume that the utility function $U:\mathbb{R}_+\to \mathbb{R}$ is $C^2$, strictly increasing, and strictly concave.%\footnote{For CRRA-type utilities defined by $U(c)=\frac{c^{1-\gamma}}{1-\gamma}, \gamma>0, \gamma\ne1$ with the coefficient of relative risk aversion $\gamma$, one may further impose Inada conditions $U'(0^+)=\infty$ and $U'(\infty)=0$, which ensure interior solutions.}

In the real world, investors rarely face frictionless markets. Portfolio allocations are often subject to regulatory, institutional, or behavioral limits that restrict borrowing, short selling, or excessive consumption. We therefore introduce a set of portfolio and consumption constraints that define the feasible control region for $(\boldsymbol{\pi}_t, C_t)$.

\vspace{0.3cm}
\noindent\emph{No Borrowing and No Short Sales.} 
The no--short-sale constraint prohibits negative holdings of risky assets, requiring that each portfolio weight satisfies $\pi_{i,t} \geq 0$ for all $i = 1, \dots, n$. 
Combined with the wealth allocation identity 
\[
\pi_{0,t} = 1 - \sum_{j=1}^n \pi_{j,t},
\]
the additional condition $\pi_{0,t} \geq 0$ rules out borrowing against the risk-free asset. 
Together, these constraints ensure that the portfolio is both unlevered and long-only. 
Such restrictions are standard in the management of regulated investment funds and pension portfolios, where leverage and short positions are typically prohibited by policy or regulation.

\vspace{0.2cm}
\noindent\emph{Consumption Bounds.} 
Consumption is constrained to remain within an admissible interval,
\[
C_{\min} \leq C_t \leq C_{\max},
\]
where $C_{\min} \geq 0$ guarantees nonnegative spending, and $C_{\max}$ represents an upper bound that may reflect either an exogenous liquidity constraint or a wealth-proportional cap, such as 
\[
C_t \leq m X_t, \quad m \in (0,1).
\]
This box-type formulation is sufficiently flexible to accommodate both fixed and state-dependent limits, capturing a wide range of practical consumption constraints encountered in portfolio choice and intertemporal optimization problems.

\vspace{0.3cm}
\noindent\emph{Unified Constraint Representation.} 
For convenience, we summarize all portfolio and consumption constraints introduced above by means of a single vector-valued inequality, 
\[
\Gamma(\boldsymbol{\pi}_t, C_t) \geq 0, \quad \forall\, t \in [0, T].
\]
Specifically, the no-borrowing, no–short-sale, and consumption-bound constraints can be written as 
\[
\Gamma(\boldsymbol{\pi}_t, C_t) = \big( \pi_{0,t},\, \pi_{1,t},\, \dots,\, \pi_{n,t},\, C_t - C_{\min},\, C_{\max} - C_t \big).
\]
Thus, the nonnegativity of portfolio weights and the lower and upper bounds on consumption are expressed in a unified notation. 

At each time–state pair $(t,x)$, the feasible control set is given by
\[
\mathcal{U}(t,x)
= \Bigl\{\,(\boldsymbol{\pi},\,C)\in \mathbb{R}^{n+1}\times\mathbb{R}_+ \;:\;
\mathbf{1}^\top \boldsymbol{\pi}=1,\ \ \Gamma\!\bigl(t,x;\boldsymbol{\pi},C\bigr)\ge \boldsymbol{0}\,\Bigr\}.
\]
A control process $(\boldsymbol{\pi}_t, C_t)$ is said to be \emph{admissible} if it is progressively measurable with respect to the filtration $\{\mathcal{F}_t\}$, satisfies $(\boldsymbol{\pi}_t, C_t) \in \mathcal{U}(t, X_t)$ almost surely for all $t \in [0, T]$, and the associated wealth SDE admits a unique strong solution $(X_t)_{t\in[0,T]}$ such that
\[
\mathbb{E}\!\int_0^T \big(\|\boldsymbol{\pi}_t\|^2 + C_t^2\big)\,dt < \infty,\quad
X_t>0\ \text{a.s. for all }t\in[0,T],\quad
\mathbb{E}\!\Big[\sup_{0\le t\le T} X_t^2\Big]<\infty .
\]

%The admissible control provides the natural entry point for the dynamic programming (DP) approach, where admissibility appears explicitly through the constrained control domain $\mathcal{U}(t,x)$ in the Hamilton-Jacobi-Bellman (HJB) equation. 

\subsection{Existing Approaches and Their Limitations}\label{sec:value_limitations}

To characterize the investor’s optimal decisions, we define the value function $V(t,x)$ as the maximal expected utility attainable starting at time $t$ with current wealth $x$. 
Given the admissible control set $\mathcal{U}(t,x)$, the investor’s optimization problem is formulated as
\begin{equation}\label{eq:value_function}
V(t,x) 
= 
\sup_{(\boldsymbol{\pi}_s,\, C_s) \in \mathcal{U}(s, X_s)} 
\mathbb{E}\left[
\int_t^T e^{-\rho s}\, U(C_s)\, ds 
+ K e^{-\rho T}\, U(X_T) 
\;\Big|\; X_t = x
\right],
\end{equation}
where $U(\cdot)$ is the instantaneous utility function, $\rho > 0$ is the subjective discount rate, and $K > 0$ represents the weight placed on terminal wealth utility.

The constrained portfolio–consumption problem described above has been extensively studied under both the dynamic programming (DP) and dual formulations. 
According to the dynamic programming principle, the value function $V$ satisfies the Hamilton–Jacobi–Bellman (HJB) equation
\begin{equation}\label{eq:HJB_constrained}
0 = \sup_{(\boldsymbol{\pi},\,C) \in \mathcal{U}(t,x)} 
\left\{
e^{-\rho t} U(C) 
+ V_t 
+ \big[ x\,\boldsymbol{\pi}^\top \widetilde{\boldsymbol{\mu}} - C \big] V_x 
+ \tfrac{1}{2} x^2 \big( \boldsymbol{\pi}^\top \widetilde{\boldsymbol{\Sigma}} \boldsymbol{\pi} \big) V_{xx} 
\right\},
\end{equation}
where the supremum is taken over $\mathcal{U}(t,x)$; 
$V_t:=\partial_t V$, $V_x:=\partial_x V$, and $V_{xx}:=\partial_{xx}V$; 
and $\widetilde{\boldsymbol{\Sigma}}:=\widetilde{\boldsymbol{v}}\,\widetilde{\boldsymbol{v}}^\top$.

It is useful to distinguish between the functional $J(\boldsymbol{\pi}, C)$, which measures the expected utility associated with a given admissible policy $(\boldsymbol{\pi}, C)$, and the value function $V(t,x)$, which represents the maximal expected utility attainable from the initial state $(t,x)$. 
In the absence of portfolio or consumption constraints, the HJB equation \eqref{eq:HJB_constrained} admits the classical closed-form Merton solutions \citep{merton1969lifetime, merton1971optimum}:
\[
C_t^* = (U')^{-1}\!\left( e^{\rho t} V_x \right), 
\quad 
\boldsymbol{\pi}_{1:n,t}^* 
= -\,\frac{V_x}{x V_{xx}}\, \boldsymbol{\Sigma}^{-1} (\boldsymbol{\mu} - r\mathbf{1}),
\quad
\pi_{0,t}^* = 1 - \mathbf{1}^\top \boldsymbol{\pi}_{1:n,t}^* .
\]

\noindent
To ensure the well-posedness of the constrained optimization problem, we impose regularity and qualification conditions on the inequality constraint functions $\Gamma_j(t,x;\boldsymbol{\pi}, C) \ge 0$ as follows.

\begin{assumption}[Constraint Regularity and Qualification]\label{ass:Gamma}
The constraint functions $\Gamma_j(t,x;\boldsymbol{\pi}, C) \ge 0$ satisfy the following conditions on the working domain:
\begin{enumerate}[label=(\roman*), leftmargin=*]

\item \emph{Smoothness and bounded derivatives:}  
Each $\Gamma_j$ is twice continuously differentiable ($C^2$) in $(\boldsymbol{\pi}, C)$ and continuously differentiable ($C^1$) in $(t, x)$, with all partial derivatives bounded on compact subsets.

\item \emph{Interior feasibility (Slater condition):}  
For every $(t, x)$, there exists a strictly feasible control $(\bar{\boldsymbol{\pi}}, \bar C)$ such that $\Gamma_j(t, x; \bar{\boldsymbol{\pi}}, \bar C) > 0$ for all $j$.

\item \emph{Constraint qualification (LICQ):}
At any maximizer or barrier stationary point, the set of active constraint gradients $\{\nabla_{(\boldsymbol{\pi}, C)} \Gamma_j\}_{j \in \mathcal{A}}$ together with the gradient of the portfolio weight constraint $\nabla_{\boldsymbol{\pi}} (\mathbf{1}^\top \boldsymbol{\pi} - 1) (= \mathbf{1})$ are linearly independent.
\end{enumerate}
\end{assumption}

\noindent
These regularity conditions encompass most linear constraints commonly used in practice, including nonnegativity of portfolio weights, sector exposure limits of the form $A\boldsymbol{\pi} \leq b$, and box-type bounds on consumption. Such constraints are standard in interior-point methods and Karush–Kuhn–Tucker (KKT) analyses \citep{boyd2004convex, nesterov1994interior, bertsekas1997nonlinear}. 
Path-dependent restrictions, such as consumption ratcheting \citep{dybvig1995dusenberry}, expressed as 
\[
C_t \geq Y_t := \max_{s < t} C_s,
\]
can also be accommodated by augmenting the state vector with the auxiliary variable $Y_t$. However, non-differentiable or combinatorial constraints — for example, buy-in thresholds or proportional transaction costs with kinked objective functions — lie beyond the scope of the present analysis.

Once inequality constraints such as nonnegative portfolio weights or bounded consumption are imposed, the pointwise maximization in the HJB equation \eqref{eq:HJB_constrained} becomes conditional on which constraints are binding. For brevity, the constrained problem can be sketched in a variational–inequality form:
\begin{equation}\label{eq:hjb_vi}
\min\!\Big\{
- V_t 
- \sup_{(\boldsymbol{\pi},C)\in\,\mathbb{R}^{n+1}\times\mathbb{R}}
\big[ e^{-\rho t} U(C) 
+ ( x\,\boldsymbol{\pi}^\top \widetilde{\boldsymbol{\mu}} - C ) V_x 
+ \tfrac{1}{2} x^2 \boldsymbol{\pi}^\top \widetilde{\boldsymbol{\Sigma}} \boldsymbol{\pi}\, V_{xx} 
\big],\ 
\min_j \Gamma_j\big(t,x;\hat{\boldsymbol{\pi}}, \hat C\big)
\Big\}=0,
\end{equation}
where $(\hat{\boldsymbol{\pi}},\hat C)$ denotes any pointwise maximizer of the 
bracketed objective at $(t,x)$, and 
$\Gamma(\cdot)\ge0$ encodes admissibility together with the 
full-investment equality $\mathbf 1^\top\boldsymbol{\pi}=1$. 
The full KKT complementarity representation is given in Section~\ref{sec:PMP_constrained}.

The variational inequality \eqref{eq:hjb_vi} constitutes a nonlinear free-boundary problem, in which both the value function $V$ and the endogenous switching surfaces that separate regions of active and inactive constraints must be determined simultaneously \citep{karatzas1987optimal, cvitanic1993hedging, XuShreve1992a, XuShreve1992b, fleming2006controlled}. 

In low-dimensional settings, numerical solutions can be obtained using finite-difference or finite-element schemes on discretized state grids. However, the computational cost of such grid-based methods grows exponentially with the number of risky assets — a manifestation of the \emph{curse of dimensionality} — making them infeasible for realistically sized portfolios. Furthermore, the nonsmooth behavior of the value function near switching boundaries often induces numerical instability, even in moderate-dimensional cases.

An alternative to the direct HJB formulation is to incorporate portfolio and consumption constraints through \emph{Lagrange multipliers}, which are often interpreted as \emph{shadow prices}. Each inequality constraint $\Gamma_j(t, x; \boldsymbol{\pi}_t, C_t) \geq 0$ is associated with a nonnegative multiplier $\nu_{j,t} \geq 0$, leading to an augmented Hamiltonian
\[
\widetilde{\mathcal H}(t, x, \boldsymbol{\pi}_t, C_t, \lambda_t, \mathbf{Z}_t; \nu_t)
= \mathcal H(t, x, \boldsymbol{\pi}_t, C_t, \lambda_t, \mathbf{Z}_t)
+ \sum_{j=1}^m \nu_t^j\, \Gamma_j(t, x; \boldsymbol{\pi}_t, C_t),
\]
where the multipliers satisfy the \emph{complementary-slackness condition} 
\[
\nu_{j,t}\,\Gamma_j(t, x; \boldsymbol{\pi}_t, C_t) = 0.
\]
The shadow prices $\nu_{j,t}$ quantify the marginal utility loss — or equivalently, the marginal cost — of tightening each constraint, thereby offering a unified economic interpretation of feasibility.

For instance, the nonnegativity constraint on the money-market position, $\pi_{0,t} \geq 0$, can be interpreted as a \emph{borrowing-rate wedge} $r^* - r$, which raises the effective financing cost whenever the constraint binds. The shadow rate $r^*$ is not an exogenous parameter but an endogenous outcome determined by the Lagrange multiplier \citep[see][]{karatzas1998methods}. Likewise, under short-sale restrictions $\pi_{i,t} \geq 0$ for $i=1,\dots,n$, the multipliers $\nu_{j,t}$ represent implicit costs of holding negative positions.\footnote{In the classical CRRA (constant relative risk aversion) setting with constant $(\boldsymbol{\mu},\boldsymbol{\Sigma})$, where $U(C)= \frac{C^{1-\gamma}}{1-\gamma}$ when $\gamma\ne 1$ and $U(C)=\log C$ when $\gamma=1$, fixing an active set 
$\mathcal{A}$ (the indices where $\pi_i > 0$), the KKT system with a full-investment multiplier $\eta \geq 0$ yields the following \emph{piecewise closed-form} solution:
\[
\boldsymbol{\pi}_{\mathcal A}
= \frac{1}{\gamma}\, \boldsymbol{\Sigma}_{\mathcal A \mathcal A}^{-1}
\bigl( (\boldsymbol{\mu} - r\mathbf{1})_{\mathcal A} - \eta\,\mathbf{1}_{\mathcal A} \bigr), 
\qquad
\boldsymbol{\pi}_{\mathcal A^c} = 0,
\]
with either $\mathbf{1}_{\mathcal A}^\top \boldsymbol{\pi}_{\mathcal A} < 1$ (hence $\eta = 0$) or, if $\mathbf{1}_{\mathcal A}^\top \boldsymbol{\pi}_{\mathcal A} = 1$,
\[
\eta = 
\frac{\mathbf{1}_{\mathcal A}^\top \boldsymbol{\Sigma}_{\mathcal A \mathcal A}^{-1} (\boldsymbol{\mu} - r\mathbf{1})_{\mathcal A} - \gamma}
{\mathbf{1}_{\mathcal A}^\top \boldsymbol{\Sigma}_{\mathcal A \mathcal A}^{-1} \mathbf{1}_{\mathcal A}}.
\]
The active set $\mathcal{A}$ is determined by the KKT/complementarity conditions and is found by solving the associated strictly convex quadratic program (QP) using an active-set pruning procedure. See, e.g., \citet{karatzas1987optimal, cvitanic1993hedging, XuShreve1992a, XuShreve1992b}.}

For consumption constraints of the form $C_{\min} \leq C_t \leq C_{\max}$, the associated multipliers define upper and lower \emph{shadow wedges} that determine the optimal consumption path. In particular, the optimal consumption policy is given by
\[
C_t^* = \min\left\{ \max\left\{ (U')^{-1}(e^{\rho t}\lambda_t),\, C_{\min} \right\},\, C_{\max} \right\}
\]
\citep[see][]{fleming2006controlled}.

In general, however, solving for the shadow prices themselves constitutes a nonlinear free-boundary problem \citep{XuShreve1992a, XuShreve1992b, fleming2006controlled}, since the active set of constraints changes endogenously with the state variables. As a result, the dual or shadow-price formulation is rarely tractable in high-dimensional settings, despite its appealing economic interpretation.

Recent research has sought to overcome these computational challenges by employing neural-network-based approximations. Two notable approaches are the \emph{Deep BSDE} methods \citep{han2018solving, e2017deep}, which approximate the pair $(V, V_x)$ by solving a system of coupled forward–backward SDEs, and \emph{Physics-Informed Neural Networks} (PINNs) \citep{raissi2019physics}, which learn the solution to the HJB equation by minimizing its residual loss.

In the absence of constraints, the value function and its gradient can be characterized by a standard forward–backward SDE (FBSDE) system \citep{pardouxpeng1990adapted}. The Deep BSDE approach approximates the pair $(\Lambda_t, \mathbf{Z}_t)$ in
\[
\begin{aligned}
dX_t &= \big( X_t \boldsymbol{\pi}_t^\top \widetilde{\boldsymbol{\mu}} - C_t \big)\, dt 
      + X_t \boldsymbol{\pi}_t^\top \widetilde{\boldsymbol{v}}\, d\mathbf{W}_t, \\
d\Lambda_t &= -f(t, X_t, \boldsymbol{\pi}_t, C_t, \Lambda_t, \mathbf{Z}_t)\, dt 
             + \mathbf{Z}_t^\top d\mathbf{W}_t, 
\qquad \Lambda_T = K e^{-\rho T} U'(X_T) %\Lambda_T = U'(X_T),
\end{aligned}
\]
where $\Lambda_t$ represents the adjoint process, i.e., the marginal value of wealth, and $\mathbf{Z}_t$ captures the sensitivity of $\Lambda_t$ to stochastic shocks.

When inequality constraints are present, however, the BSDE must incorporate an additional 
\emph{reflection term} to enforce feasibility:
\[
d\Lambda_t = -f(\cdot)\, dt + \mathbf{Z}_t^\top d\mathbf{W}_t + dK_t,\qquad 
K_t=(K_t^{\,1},\dots,K_t^{\,m}),\ K^{\,j}_0=0,
\]
with each component $K^{\,j}$ nondecreasing, and the Skorokhod conditions
\[
\Gamma_j\big(t,X_t;\boldsymbol{\pi}_t, C_t\big)\ge 0,\qquad
\int_0^T \Gamma_j\big(t,X_t;\boldsymbol{\pi}_t, C_t\big)\, dK_t^{\,j} = 0,\quad j=1,\dots,m.
\]
This is precisely a \emph{reflected BSDE} (RBSDE): the process $K$ keeps the solution in the feasible set and 
increases only when the corresponding constraint is active.

Numerically, solving the reflected problem is significantly more challenging: the neural network must approximate not only the pair $(\Lambda_t, \mathbf{Z}_t)$ but also the \emph{free-boundary structure} encoded in $K_t$, whose dynamics can become highly irregular and unstable in high-dimensional settings.\footnote{
For the unconstrained BSDE, existence and uniqueness are guaranteed under standard Lipschitz continuity and linear-growth conditions on the driver, together with square-integrable terminal data \citep{pardouxpeng1990adapted}. 
For the reflected case, if the barrier and driver satisfy the usual regularity and compatibility conditions (e.g., Mokobodzki-type), there exists a unique triple $(\Lambda, \mathbf{Z}, K)$ solving the RBSDE \citep{el1997reflected}. 
In our setting, the bounded and Lipschitz coefficients, together with the feasibility structure implied by Assumption~\ref{ass:Gamma}, ensure that these (R)BSDE well-posedness results apply at the modeling level considered here.
}

While highly effective for smooth, unconstrained problems, these value-based neural methods encounter fundamental difficulties in the presence of inequality constraints, since the associated obstacle conditions must be learned implicitly rather than enforced directly. PINNs tend to exhibit large residual variance near nonsmooth regions of the HJB solution, while Deep BSDE approaches struggle with reflection terms that can behave discontinuously when constraints switch on and off along simulated paths. Consequently, existing neural PDE and BSDE solvers often fail to achieve reliable accuracy in constrained, high-dimensional portfolio problems.

In summary, the dynamic programming (DP), dual (shadow-price), and neural value-based approaches all provide rigorous characterizations of the optimal solution but suffer from significant computational limitations once realistic constraints and multi-asset settings are introduced. Specifically, grid-based HJB solvers are severely limited by the curse of dimensionality; dual formulations require solving nonlinear complementarity systems whose structure changes endogenously; and value-based neural methods remain fragile in regions where constraints become binding. These challenges motivate a \emph{policy-centric} alternative that focuses directly on the control variables themselves rather than on the value function, thereby bypassing many of the computational bottlenecks inherent in traditional formulations.

\section{Pontryagin's Maximum Principle Approach to Constrained Multi-Asset Portfolio Optimization}
\label{sec:PMP_multiasset}

\noindent
This section develops a framework based on \emph{Pontryagin’s Maximum Principle} (PMP), which provides first-order necessary conditions for optimality in continuous time without requiring an explicit value function or a dynamic-programming grid. The PMP formulation characterizes optimal portfolio and consumption strategies through a coupled system of forward–backward stochastic differential equations (FBSDEs) that link the state dynamics with their corresponding adjoint (costate) processes.

\subsection{Unconstrained PMP Formulation}
\label{sec:PMP_unconstrained}

\noindent
We begin with the multi-asset Merton problem introduced in Section~\ref{sec:dp_multiasset}, allowing borrowing and short-selling and imposing no explicit consumption constraints. Under Pontryagin’s Maximum Principle (PMP), the optimal solution is characterized by a system of forward–backward stochastic differential equations (FBSDEs).\footnote{The FBSDE framework originates from \citet{pontryagin2018mathematical} and has been extensively developed in subsequent works such as \citet{pardouxpeng1990adapted}, \citet{fleming2006controlled}, and \citet{pham2009continuous}.} 

Central to this approach is the introduction of \emph{adjoint processes}, consisting of a scalar adjoint $\lambda_t$ associated with the state variable $X_t$ and a vector adjoint $\mathbf{Z}_t \in \mathbb{R}^n$ corresponding to the $n$-dimensional Brownian motion $\mathbf{W}_t$ driving asset returns. The process $\lambda_t$ represents the \emph{marginal value of wealth}, directly analogous to the partial derivative $V_x(t, X_t)$ of the value function in the dynamic programming formulation.

Under the standard smoothness and integrability conditions, Pontryagin’s Maximum Principle defines the Hamiltonian
\begin{equation}
\label{eq:hamiltonian_unconstrained}
\mathcal{H}\bigl(t, X_t, \boldsymbol{\pi}_t, C_t, \lambda_t, \mathbf{Z}_t\bigr)
=
e^{-\rho t} U(C_t)
+ \lambda_t \bigl[ X_t\, \boldsymbol{\pi}_t^\top \widetilde{\boldsymbol{\mu}} - C_t \bigr]
+ \mathbf{Z}_t^\top \bigl( X_t\, \widetilde{\boldsymbol{v}}^\top \boldsymbol{\pi}_t \bigr),
\end{equation}
and the associated coupled forward–backward system characterizing the optimal state and adjoint dynamics:
\begin{align}
dX_t^*
&= \Big[ X_t^* (\boldsymbol{\pi}_t^*)^\top \widetilde{\boldsymbol{\mu}} - C_t^* \Big]\, dt
+ X_t^* (\boldsymbol{\pi}_t^*)^\top \widetilde{\boldsymbol{v}}\, d\mathbf{W}_t,
\quad X_0^* = x_0, \label{eq:pmp_state} \\[0.5em]
d\lambda_t^*
&= -\,\partial_x \mathcal{H} \bigl( t, X_t^*, \boldsymbol{\pi}_t^*, C_t^*, \lambda_t^*, \mathbf{Z}_t^* \bigr)\, dt
+ \mathbf{Z}_t^{*\top}\, d\mathbf{W}_t,
\quad \lambda_T^* = K e^{-\rho T} U'(X_T^*). \label{eq:pmp_adjoint}
\end{align}

Optimal controls maximize the Hamiltonian pointwise at each time–state pair:
\[
(\boldsymbol{\pi}_t^*, C_t^*) 
= \arg\max_{\boldsymbol{\pi},\, C}\;
\mathcal{H}(t, X_t, \boldsymbol{\pi}, C, \lambda_t, \mathbf{Z}_t).
\]
The first-order (stationarity) conditions are then given by
\begin{equation}\label{eq:pmp_foc_c}
\frac{\partial \mathcal{H}}{\partial C}
= e^{-\rho t} U'(C_t^*) - \lambda_t^* = 0,
\end{equation}
\begin{equation}\label{eq:pmp_foc_pi}
\frac{\partial \mathcal{H}}{\partial \boldsymbol{\pi}}
= X_t^* \lambda_t^*\, \widetilde{\boldsymbol{\mu}}
+ X_t^* \widetilde{\boldsymbol{v}}\, \mathbf{Z}_t^* = 0.
\end{equation}

From \eqref{eq:pmp_foc_c}, the optimal consumption policy is obtained as
\begin{equation}\label{eq:pmp_C_explicit_general}
C_t^* = \bigl(U'\bigr)^{-1} \!\bigl( e^{\rho t} \lambda_t^* \bigr),
\end{equation}
and substituting \eqref{eq:pmp_foc_pi} yields the optimal risky asset allocation:
\begin{equation}\label{eq:pmp_pi_explicit_general}
\boldsymbol{\pi}_{1:n,t}^*
=
-\,\frac{\lambda_t^*}{X_t^* \, (\partial_x \lambda_t^*)}\, 
\boldsymbol{\Sigma}^{-1} \bigl( \boldsymbol{\mu} - r \mathbf{1} \bigr),
\end{equation}
which coincides with the familiar Merton portfolio rule when $\lambda_t = V_x(t, X_t)$ and $\partial_x \lambda_t = V_{xx}(t, X_t)$.

Thus, the unconstrained PMP formulation reproduces the structure of the dynamic programming solution but does so \emph{without} requiring explicit computation of the value function.\footnote{
If one imposes the full-investment constraint $\mathbf{1}^\top \boldsymbol{\pi}_t = 1$, a Lagrange multiplier $\eta_t$ can be introduced into $\mathcal{H}$. Eliminating $\eta_t$ leads to \eqref{eq:pmp_pi_explicit_general} for the risky block. Under mild regularity conditions, the relation $\lambda_t = V_x(t, X_t)$ implies $\partial_x \lambda_t = V_{xx}(t, X_t)$, ensuring that \eqref{eq:pmp_pi_explicit_general} matches the dynamic-programming solution.}
This Hamiltonian formulation provides a natural foundation for incorporating inequality constraints in subsequent sections.

To ensure that the FBSDE system \eqref{eq:pmp_state}--\eqref{eq:pmp_adjoint} is mathematically well-posed and that the PMP characterization remains valid, we impose standard smoothness and integrability requirements.

\begin{assumption}[PMP Regularity for the Unconstrained Case]\label{ass:pmp_uncon}
The following conditions hold:
\begin{enumerate}[label=(\roman*), leftmargin=*]

\item \emph{Hamiltonian differentiability.}  
The Hamiltonian $\mathcal{H}$ is continuously differentiable ($C^1$) in $(x, \boldsymbol{\pi}, C)$, and its partial derivatives $\partial_x \mathcal{H}$, $\nabla_{\boldsymbol{\pi}} \mathcal{H}$, and $\partial_C \mathcal{H}$ are locally Lipschitz (properties that follow from the model and utility assumptions). Moreover, for sufficiency, we require that $(\boldsymbol{\pi}, C) \mapsto \mathcal{H}$ be concave (which is immediate here, since $\mathcal{H}$ is linear in $\boldsymbol{\pi}$ and concave in $C$), ensuring that the pointwise maximizers solve the control block.

\item \emph{FBSDE well-posedness.}  
The coupled system \eqref{eq:pmp_state}--\eqref{eq:pmp_adjoint} admits a unique $\mathcal{F}$-adapted solution under standard Lipschitz and monotonicity conditions on the coefficients and on the driver $-\partial_x \mathcal{H}$ \citep[see, e.g.,][]{yong2012stochastic, ma1999forward}.

%\item \emph{Risky weights.}  
%For the explicit formula \eqref{eq:pmp_pi_explicit_general}, we assume that $\lambda_t$ is differentiable in $x$ with %$\partial_x \lambda_t$ almost surely nonzero (typically $\partial_x \lambda_t < 0$ on compact domains), and that %$\boldsymbol{\Sigma}^{-1}$ exists almost surely.
\end{enumerate}
\end{assumption}

\noindent
In Assumption~\ref{ass:pmp_uncon}, condition~(i) justifies the differentiation of the Hamiltonian and guarantees that the stationarity conditions provide both necessary and sufficient conditions for optimality. Condition~(ii) ensures the existence and uniqueness of the forward–backward system, thereby making the PMP formulation mathematically meaningful. Condition~(iii) is a technical strengthening that is only required for obtaining the explicit portfolio solution but not for the general PMP characterization itself. Under Assumption~\ref{ass:pmp_uncon}, the unconstrained PMP problem is well-defined, and the optimal control obtained from the first-order conditions coincides with the solution from the dynamic programming formulation. This establishes a solid theoretical foundation for extending the PMP framework to portfolio problems with inequality constraints.

\subsection{Constrained PMP with Barrier Regularization}
\label{sec:PMP_constrained}

When the admissible control set $\mathcal{U}(t,x)$ is restricted by inequality constraints $\Gamma_j(\boldsymbol{\pi}_t, C_t) \ge 0$, the unconstrained first-order conditions \eqref{eq:pmp_foc_c}--\eqref{eq:pmp_foc_pi} no longer apply directly, as the feasible region may contain non-smooth boundaries where differentiability breaks down. To incorporate such constraints while preserving differentiability, we introduce a \emph{barrier-regularized Hamiltonian}, which penalizes the approach to the boundary of the admissible set and thereby enforces feasibility through an interior-point formulation \citep{nesterov1994interior, boyd2004convex, wright1997primal}.

At each time–state pair $(t, x)$, the barrier-regularized stagewise optimization problem is formulated as
\begin{equation}\label{eq:H_bar_again}
\max_{(\boldsymbol{\pi}, C) \in \mathcal U_{\mathrm{int}}(t,x)} \widetilde{\mathcal H}_{\mathrm{bar}}
:=
\max_{(\boldsymbol{\pi}, C) \in \mathcal U_{\mathrm{int}}(t,x)} 
\left\{
\mathcal{H}
+ \eta_t \bigl( 1 - \mathbf{1}^\top \boldsymbol{\pi}_t \bigr)
+ \varepsilon_{\mathrm{bar}} \sum_{j=1}^m \ln \Gamma_j(\boldsymbol{\pi}_t, C_t)
\right\},
\quad \varepsilon_{\mathrm{bar}} > 0,
\end{equation}
where we take the admissible set to be
\[
\mathcal U_{\mathrm{int}}(t,x)
= \Bigl\{\,(\boldsymbol{\pi},\,C)\in \mathbb{R}^{n+1}\times\mathbb{R}_+ \;:\;
\Gamma\!\bigl(t,x;\boldsymbol{\pi},C\bigr)>\boldsymbol{0}\,\Bigr\}.
\]
Here, $\eta_t$ denotes the Lagrange multiplier associated with the full-investment constraint $\mathbf{1}^\top \boldsymbol{\pi}_t = 1$, and $\varepsilon_{\mathrm{bar}}>0$ is a small barrier parameter that controls the distance of the optimizer from the boundary of the feasible region.

The logarithmic barrier term $\ln \Gamma_j(\boldsymbol{\pi}_t, C_t)$ diverges to $-\infty$ as any $\Gamma_j(\boldsymbol{\pi}_t, C_t) \to 0^+$, thereby ensuring that the solution remains strictly in the interior of the admissible set. This interior-point formulation guarantees that the Hamiltonian and all its derivatives remain smooth and well-defined, while the inequality constraints are enforced in the limit as $\varepsilon_{\mathrm{bar}} \to 0$.

The barrier formulation can be interpreted as a smooth approximation to the standard Karush–Kuhn–Tucker (KKT) system. Let $\nu_t = (\nu_{1,t}, \dots, \nu_{m,t})^\top$ denote the Lagrange multipliers associated with the inequality constraints $\Gamma_j(\boldsymbol{\pi}, C) \ge 0$. The conventional constrained Hamiltonian can then be written as
\begin{equation}\label{eq:H_kkt_again}
\widetilde{\mathcal H}_{\mathrm{KKT}}
=
\mathcal H
+ \eta_t \bigl( 1 - \mathbf{1}^\top \boldsymbol{\pi}_t \bigr)
+ \sum_{j=1}^m \nu_{j,t}\, \Gamma_j(\boldsymbol{\pi}_t, C_t),
\quad \nu_{j,t} \ge 0,
\end{equation}
subject to the complementary-slackness conditions $\nu_{j,t} \Gamma_j(\boldsymbol{\pi}_t, C_t) = 0$ for all $j$.
By defining the implicit mapping 
\[
\nu_{j,t}=\frac{\varepsilon_{\mathrm{bar}}}{\Gamma_j(\boldsymbol{\pi}_t,C_t)},
\]
the \emph{stationarity equations} and the \emph{equality constraint} in \eqref{eq:H_bar_again} 
coincide \emph{exactly} with those of the KKT system \eqref{eq:H_kkt_again}, 
while the complementary–slackness condition is replaced by the \emph{central-path relation} 
\(\nu_{j,t}\,\Gamma_j(\boldsymbol{\pi}_t,C_t)=\varepsilon_{\mathrm{bar}}\).
As \(\varepsilon_{\mathrm{bar}}\to0\), the barrier stationary point 
\((\hat{\boldsymbol{\pi}}_{\varepsilon_{\mathrm{bar}}},\hat C_{\varepsilon_{\mathrm{bar}}})\) 
converges to the KKT/PMP maximizer \((\boldsymbol{\pi}^*,C^*)\) with variable error 
\(\|(\hat{\boldsymbol{\pi}}_{\varepsilon_{\mathrm{bar}}},\hat C_{\varepsilon_{\mathrm{bar}}})-(\boldsymbol{\pi}^*,C^*)\|=O(\varepsilon_{\mathrm{bar}})\).

This convergence relationship parallels the \emph{central path} property of interior-point methods in static optimization, but here it operates \emph{stagewise} within the continuous-time Hamiltonian system, providing a smooth path from interior feasible controls to the optimal constrained solution.

In the specific context of portfolio and consumption controls, the feasibility mechanism introduced by the barrier-regularized PMP admits clear economic and mathematical interpretations. For non-negativity or simplex constraints on risky-asset weights $(\pi_i \geq 0,\, \mathbf{1}^\top \boldsymbol{\pi} = 1)$, the logarithmic barrier introduces curvature proportional to $\varepsilon_{\mathrm{bar}} / \pi_i^2$, which increases sharply as any $\pi_i \to 0$. This ensures that all portfolio weights remain strictly positive for any finite $\varepsilon_{\mathrm{bar}}$, while the full-investment identity is exactly enforced through the multiplier $\eta_t$. As the barrier parameter $\varepsilon_{\mathrm{bar}}$ decreases, these strictly interior weights converge to the true constrained optimum, at which inactive positions satisfy $\pi_i^* = 0$.

For box constraints $C_{\min} \leq C_t \leq C_{\max}$ (or wealth-proportional caps $0 \leq C_t \leq m X_t$), the barrier-modified first-order condition takes the form
\[
e^{-\rho t} U'(C) - \lambda_t 
+ \frac{\varepsilon_{\mathrm{bar}}}{C - C_{\min}} 
- \frac{\varepsilon_{\mathrm{bar}}}{C_{\max} - C} = 0,
\]
which admits a unique interior solution $C_t^{\varepsilon_{\mathrm{bar}}} \in (C_{\min}, C_{\max})$. In the limit $\varepsilon_{\mathrm{bar}} \to 0$, this interior solution converges to the standard KKT \emph{clipping rule}:
\[
C_t^* 
= \min\!\Bigl\{ \max\bigl\{ (U')^{-1}(e^{\rho t} \lambda_t),\, C_{\min} \bigr\},\, C_{\max} \Bigr\},
\]
thereby recovering exact feasibility and satisfying the complementary-slackness conditions.

To ensure that the barrier-regularized PMP formulation is mathematically well-posed and convergent, we impose the following stagewise regularity conditions.

\begin{assumption}[Barrier/KKT Stagewise Regularity]\label{ass:barrier_kkt}
We work under the model assumptions and the constraint qualification (LICQ) stated in Assumption~\ref{ass:Gamma}. In addition, for each $(t, x)$ we assume:
\begin{enumerate}[label=(\roman*), leftmargin=*]

\item \emph{Bounded level sets for the barrier objective:}
For any fixed $\varepsilon_{\mathrm{bar}} > 0$, the superlevel sets of $\widetilde{\mathcal H}_{\mathrm{bar}}$ are bounded in $(\boldsymbol{\pi}, C)$. \emph{(Slater’s condition is already assumed in Assumption~\ref{ass:Gamma}(ii).)}

\item \emph{KKT nondegeneracy / strong metric regularity at the solution:} LICQ holds and strict complementarity holds at $(\boldsymbol{\pi}^*,C^*,\eta^*,\nu^*)$, and the bordered KKT Jacobian of
\[
(\boldsymbol{\pi},C,\eta,\nu)\mapsto
\begin{pmatrix}
\nabla_{\boldsymbol{\pi}}\mathcal H - \eta\,\mathbf 1 + J_{\boldsymbol{\pi}}\Gamma\,\nu\\
\partial_C\mathcal H + \partial_C\Gamma^{\!\top}\nu\\
\mathbf 1^\top\boldsymbol\pi - 1\\
\mathrm{diag}(\Gamma)\,\nu
\end{pmatrix}
\]
is nonsingular at $(\boldsymbol{\pi}^*,C^*,\eta^*,\nu^*)$. \emph{(In particular, no strict concavity in $\boldsymbol{\pi}$ is assumed; for the barrier subproblem, strict concavity is provided by the log-barrier term.)}

\item \emph{Smoothness near the solution and central-path interiority:}  
The gradients $\nabla_{(\boldsymbol{\pi}, C)} \mathcal H$ and $\nabla_{(\boldsymbol{\pi}, C)} \Gamma_j$ are locally Lipschitz in a neighborhood of the solution. Moreover, along the barrier iterates, the interiority condition
\[
\Gamma_j(\boldsymbol{\pi}, C) \ge c\,\varepsilon_{\mathrm{bar}}
\]
holds for some constant $c \in (0, 1)$ and all $j$.
\end{enumerate}
\end{assumption}

\noindent
Assumption~\ref{ass:barrier_kkt} guarantees the mathematical soundness of the barrier-regularized PMP formulation. Condition~(i) ensures the existence of a strictly feasible interior point and prevents the optimizer from diverging. Condition~(ii) guarantees local uniqueness and stability of the maximizer by ruling out flat directions on the constraint manifold. Condition~(iii) provides sufficient smoothness for differentiation-based arguments and ensures that the barrier trajectory remains strictly interior, thereby enabling convergence to the KKT solution as $\varepsilon_{\mathrm{bar}} \to 0$.

Formally, let $(\boldsymbol{\pi}^*, C^*)$ denote the exact constrained PMP solution satisfying the KKT conditions, and let $(\hat{\boldsymbol{\pi}}_{\varepsilon_{\mathrm{bar}}}, \hat{C}_{\varepsilon_{\mathrm{bar}}})$ denote the maximizer of the barrier-augmented Hamiltonian for a fixed $\varepsilon_{\mathrm{bar}} > 0$. Under Assumption~\ref{ass:barrier_kkt}, there exists a differentiable curve 
\[
\{(\hat{\boldsymbol{\pi}}_{\varepsilon_{\mathrm{bar}}}, \hat{C}_{\varepsilon_{\mathrm{bar}}})\}_{\varepsilon_{\mathrm{bar}} > 0},
\]
referred to as the \emph{central path}, along which the barrier solutions remain strictly feasible and converge to the true constrained optimum:
\[
(\hat{\boldsymbol{\pi}}_{\varepsilon_{\mathrm{bar}}}, \hat{C}_{\varepsilon_{\mathrm{bar}}}) \longrightarrow (\boldsymbol{\pi}^*, C^*) 
\quad \text{as } \varepsilon_{\mathrm{bar}} \to 0.
\]

Before characterizing the convergence properties, we first verify that the barrier-regularized Hamiltonian defined in \eqref{eq:H_bar_again} gives rise to a well-posed interior optimization problem at each time–state pair $(t, x)$. This verification ensures that the stagewise maximizer $(\boldsymbol{\pi}^{\varepsilon_{\mathrm{bar}}}, C^{\varepsilon_{\mathrm{bar}}})$ in the barrier formulation indeed exists, is unique under mild concavity conditions, and lies strictly within the feasible region. 
% Because the barrier term $\varepsilon_{\mathrm{bar}} \sum_j \ln \Gamma_j(\boldsymbol{\pi}, C)$ diverges to $-\infty$ as any constraint approaches the boundary, the maximization occurs strictly in the interior of the admissible region. That is, for each $j$ there exists a constant $c \in (0,1)$ such that $\Gamma_j(\boldsymbol{\pi}^{\varepsilon_{\mathrm{bar}}}, C^{\varepsilon_{\mathrm{bar}}}) \geq c\,\varepsilon_{\mathrm{bar}}$, which guarantees strict interior feasibility for all finite $\varepsilon_{\mathrm{bar}} > 0$.

The barrier-regularized optimization problem at a fixed $(t,x)$ satisfies the stationarity condition
\[
\mathbf{F}(\boldsymbol{\pi}, \eta) =
\begin{pmatrix}
\nabla_{\boldsymbol{\pi}} \widetilde{\mathcal H}_{\mathrm{bar}}(t, x; \boldsymbol{\pi}, C, \eta) \\[0.5em]
\mathbf{1}^\top \boldsymbol{\pi} - 1
\end{pmatrix}
= \mathbf{0},
\]
which can be solved in the tangent space $\{\mathbf{1}^\top d = 0\}$ using a Newton–Conjugate-Gradient (Newton–CG) method — an efficient alternative to directly inverting the full Hessian matrix \citep{nocedal2006numerical}. Writing $g := \nabla_{\boldsymbol{\pi}} \widetilde{\mathcal H}_{\mathrm{bar}}$ and denoting by $H_{\mathrm{bar}}[\cdot]$ the reduced Hessian operator, the local search direction $d$ solves
\[
\mathcal H_{\mathrm{bar}}[d] = -\,g + \eta\,\mathbf{1}, 
\qquad \mathbf{1}^\top d = 0.
\]

\noindent
For general inequality constraints $\Gamma_j(\boldsymbol{\pi}, C) > 0$, the Hessian–vector product takes the form
\begin{equation}\label{eq:hvp_general}
% (13) corrected
\mathcal H_{\mathrm{bar}}[v]
=\; \nabla^2_{\boldsymbol{\pi}\boldsymbol{\pi}} \mathcal H\, v
\;+\; \varepsilon_{\mathrm{bar}}\sum_{l=1}^{m}
\left[
\frac{1}{\Gamma_l(\boldsymbol{\pi}, C)}\, \nabla^2_{\boldsymbol{\pi}\boldsymbol{\pi}} \Gamma_l\, v
\;-\; \frac{1}{\Gamma_l(\boldsymbol{\pi}, C)^2}\, \big(\nabla_{\boldsymbol{\pi}}\Gamma_l\,\nabla_{\boldsymbol{\pi}}\Gamma_l^\top v\big)
\right].
\end{equation}
For the special case of pure nonnegativity constraints $\Gamma_i(\boldsymbol{\pi}, C) = \pi_i$, this simplifies to
\begin{equation}\label{eq:hvp_nonneg}
\mathcal H_{\mathrm{bar}}[v] = \nabla^2_{\boldsymbol{\pi}\boldsymbol{\pi}} \mathcal H\, v 
\;-\; \mathrm{diag}\!\bigl( \varepsilon_{\mathrm{bar}} / \boldsymbol{\pi}^2 \bigr) \, v.
\end{equation}

\noindent
The diagonal barrier term introduces curvature proportional to $\varepsilon_{\mathrm{bar}}/\pi_i^2$ and diverges as $\pi_i\to 0$, which ensures \emph{negative definiteness} of $\nabla^2_{\boldsymbol{\pi}\boldsymbol{\pi}}\mathcal{H}_{\mathrm{bar}}$ on the tangent space (equivalently, \emph{positive definiteness} of $-\nabla^2_{\boldsymbol{\pi}\boldsymbol{\pi}}\mathcal{H}_{\mathrm{bar}}$). Hence the stagewise maximization is strictly concave and well-conditioned for Newton–CG updates. The full-investment equality can be handled either by projecting the Newton step onto the subspace $\{\mathbf{1}^\top d=0\}$ or by applying a $1\times 1$ Schur complement reduction in $\eta$ to solve the KKT system efficiently.\footnote{The Schur complement is a standard linear-algebra technique for solving saddle-point KKT systems efficiently \citep{nocedal2006numerical, boyd2004convex}. For the Newton step of the barrier problem, the first-order conditions yield the block system
\[
\begin{psmallmatrix}
\mathcal H_{\mathrm{bar}} & \mathbf{1}\\
\mathbf{1}^\top & 0
\end{psmallmatrix}
\begin{psmallmatrix}
d\\ \Delta\eta
\end{psmallmatrix}
=
-\begin{psmallmatrix}
g\\ c
\end{psmallmatrix},
\quad
\text{with } g:=\nabla_{\boldsymbol{\pi}}\mathcal{H}_{\mathrm{bar}}\ \text{and}\ c:=\mathbf{1}^\top\boldsymbol{\pi}-1,
\]
where $\mathcal H_{\mathrm{bar}}$ is the (reduced) Hessian with respect to $\boldsymbol{\pi}$ and $\Delta\eta$ is the multiplier update. Eliminating $d$ gives the scalar equation
\[
\mathbf{1}^\top \mathcal H_{\mathrm{bar}}^{-1}\mathbf{1}\,\Delta\eta
= c - \mathbf{1}^\top \mathcal H_{\mathrm{bar}}^{-1} g,
\]
after which
\(
d=-\,\mathcal H_{\mathrm{bar}}^{-1}\bigl(g+\mathbf{1}\,\Delta\eta\bigr)
\)
recovers the primal step. This reduction avoids inverting the full $(n+1)\times(n+1)$ KKT matrix and improves numerical efficiency.}

To maintain feasibility and ensure monotone ascent of the barrier objective, the step size $\alpha$ is selected using a \emph{fraction-to-the-boundary} line search combined with an \emph{Armijo sufficient-increase condition} \citep{wright1997primal, wachter2006implementation}. To prevent any constraint $\Gamma_j(\boldsymbol{\pi} + \alpha d, C)$ from violating interiority, the maximal admissible step is chosen as
\[
\alpha_{\max} = \min \Bigl\{ 1,\ \min_{i:\, d_i < 0} \bigl[ -\tau\, \pi_i / d_i \bigr] \Bigr\}, 
\qquad \tau \in (0,1),
\]
so that the updated point satisfies $\pi_i + \alpha d_i \ge (1 - \tau) \pi_i > 0$. This rule guarantees strict interior feasibility at every iteration.

Starting from $\alpha_{\max}$, the step length is then decreased until the Armijo condition holds:
\[
\mathcal{H}_{\mathrm{bar}}(\boldsymbol{\pi} + \alpha d, C, \lambda_t, \mathbf{Z}_t) 
\ge 
\mathcal{H}_{\mathrm{bar}}(\boldsymbol{\pi}, C, \lambda_t, \mathbf{Z}_t) 
+ \sigma \alpha g^\top d, 
\quad \sigma \in (0,1),
\]
where $g = \nabla_{\boldsymbol{\pi}} \mathcal{H}_{\mathrm{bar}}$. This is the standard sufficient-increase criterion in interior-point Newton methods: it ensures that each accepted step yields a monotone improvement of the barrier-regularized Hamiltonian while maintaining $\Gamma_j(\boldsymbol{\pi}, C) \ge c \,\varepsilon_{\mathrm{bar}}$ with $c \in (0,1)$ for all $j$. 

The combined fraction-to-the-boundary and Armijo procedures therefore guarantee both \emph{strict interior feasibility} and \emph{monotone ascent} of the barrier-regularized objective at every stage.\footnote{The consumption optimization block is scalar. With box constraints, the closed-form clipping rule
\[
C^* = \min\bigl\{ \max\{ (U')^{-1}(e^{\rho t}\lambda_t),\, C_{\min} \},\, C_{\max} \bigr\}
\]
solves the problem exactly. Under a barrier formulation, the stationarity condition
\[
e^{-\rho t} U'(C) - \lambda_t + \frac{\varepsilon_{\mathrm{bar}}}{C - C_{\min}} - \frac{\varepsilon_{\mathrm{bar}}}{C_{\max} - C} = 0
\]
can instead be solved by a short damped Newton iteration with the same line-search logic \citep{nocedal2006numerical, boyd2004convex}.}

In sum, under Assumptions~\ref{ass:Gamma} and \ref{ass:barrier_kkt}, the barrier-regularized Hamiltonian admits a unique stationary point satisfying $\Gamma_j(\boldsymbol{\pi}^{\varepsilon_{\mathrm{bar}}}, C^{\varepsilon_{\mathrm{bar}}}) \ge c\,\varepsilon_{\mathrm{bar}}$ and ensuring positive definiteness (or strict concavity) of $\mathcal H_{\mathrm{bar}}$ on the tangent space. This stationary point defines a strictly interior optimal control for every finite $\varepsilon_{\mathrm{bar}} > 0$. Consequently, the stagewise barrier subproblem is well posed, the Newton–CG iteration is locally convergent, and the resulting sequence of barrier solutions traces a differentiable central path converging to the KKT manifold as $\varepsilon_{\mathrm{bar}} \to 0$ \citep{bertsekas1997nonlinear, nocedal2006numerical}. 

In practical implementations, only a modest number of inner Newton steps and a single backtracking line search are typically sufficient to reduce the stagewise residual to the tolerance required by the outer learning loop. There is no need to oversolve this microproblem at each stage. The resulting barrier maximizers closely follow the central path and, by Theorem~\ref{thm:barrier_policy} below, achieve an $O(\varepsilon_{\mathrm{bar}})$ policy error and an $O(\varepsilon_{\mathrm{bar}}^2)$ instantaneous Hamiltonian gap.

The following theorem summarizes the approximation properties of the barrier-regularized solution, where we suppress the explicit dependence on $(t,x)$ and the adjoint variables $(\lambda_t, \mathbf{Z}_t)$ for readability.

\begin{theorem}[Barrier vs.\ KKT policy: first-order policy accuracy, quadratic Lagrangian gap, linear Hamiltonian gap]\label{thm:barrier_policy}
Under Assumptions~\ref{ass:Gamma} and \ref{ass:barrier_kkt},  
there exist constants $K, \kappa_1, \kappa_2, C_{H,1}, C_{H,2} > 0$ (independent of $\varepsilon_{\mathrm{bar}}$) and $\bar{\varepsilon} > 0$ such that, for all $0 < \varepsilon_{\mathrm{bar}} \le \bar{\varepsilon}$, the barrier maximizer $(\hat{\boldsymbol{\pi}}_{\varepsilon_{\mathrm{bar}}}, \hat{C}_{\varepsilon_{\mathrm{bar}}})$ satisfies
\begin{align}
\|(\hat{\boldsymbol{\pi}}_{\varepsilon_{\mathrm{bar}}}, \hat{C}_{\varepsilon_{\mathrm{bar}}}) - (\boldsymbol{\pi}^*, C^*)\| 
&\le K\,\varepsilon_{\mathrm{bar}}, \label{eq:pol_err_main_revised} \\[0.5em]
0 \le \mathcal{L}(\boldsymbol{\pi}^*, C^*;\eta^*,\nu^*) - \mathcal{L}(\hat{\boldsymbol{\pi}}_{\varepsilon_{\mathrm{bar}}}, \hat{C}_{\varepsilon_{\mathrm{bar}}};\eta^*,\nu^*)
&\le \tfrac{L K^2}{2}\,\varepsilon_{\mathrm{bar}}^2, \label{eq:lag_gap_main_revised} \\[0.5em]
0 \le \mathcal H(\boldsymbol{\pi}^*, C^*) - \mathcal H(\hat{\boldsymbol{\pi}}_{\varepsilon_{\mathrm{bar}}}, \hat{C}_{\varepsilon_{\mathrm{bar}}})
&\le C_{H,1}\,\varepsilon_{\mathrm{bar}} + C_{H,2}\,\varepsilon_{\mathrm{bar}}^2, \label{eq:ham_gap_main_revised} \\[0.5em]
0 \le J(\boldsymbol{\pi}^*, C^*) - J(\hat{\boldsymbol{\pi}}_{\varepsilon_{\mathrm{bar}}}, \hat{C}_{\varepsilon_{\mathrm{bar}}})
&\le \kappa_1\,\varepsilon_{\mathrm{bar}}^2 + \kappa_2\,\varepsilon_{\mathrm{bar}}. \label{eq:value_gap_main_revised}
\end{align}
Here $\mathcal L(\cdot;\eta^*,\nu^*)=\mathcal H(\cdot)+\eta^*(1-\mathbf 1^\top\pi)+\nu^{*\top}\Gamma(\cdot)$ is the concave Lagrangian evaluated at the KKT multipliers $(\eta^*,\nu^*)$.
\end{theorem}

\begin{proof}
A complete proof is provided in Appendix~\ref{app:proof_barrier_policy}.
\end{proof}

\noindent
The proof in Appendix~\ref{app:proof_barrier_policy} follows the classical perturbation analysis of interior-point methods. Linearizing the first-order conditions around the KKT solution shows that the barrier stationary point satisfies a perturbed optimality system with residuals of order $O(\varepsilon_{\mathrm{bar}})$. This yields the $O(\varepsilon_{\mathrm{bar}})$ policy accuracy \eqref{eq:pol_err_main_revised}, an $O(\varepsilon_{\mathrm{bar}}^2)$ \emph{Lagrangian} gap at the KKT multipliers \eqref{eq:lag_gap_main_revised}, and an $O(\varepsilon_{\mathrm{bar}})$ \emph{Hamiltonian} gap with an $O(\varepsilon_{\mathrm{bar}}^2)$ remainder \eqref{eq:ham_gap_main_revised}. Thus, Theorem~\ref{thm:barrier_policy} establishes that the barrier solution converges \emph{linearly} to the true KKT solution in control variables, while the Lagrangian suboptimality decreases \emph{quadratically} and the Hamiltonian suboptimality decreases \emph{linearly} in $\varepsilon_{\mathrm{bar}}$. As a result, feasibility is preserved exactly at every stage, and the induced suboptimality diminishes at a provably controlled rate.%
\footnote{If no inequality constraint is active at the KKT solution, the first-order Hamiltonian term vanishes and the Hamiltonian gap also becomes $O(\varepsilon_{\mathrm{bar}}^2)$.}

From a practical perspective, the barrier-regularized PMP formulation offers a fully differentiable and computationally stable mechanism for enforcing feasibility at each time–state pair. In contrast to standard KKT solvers—which require explicit identification of the active set—the barrier approach keeps the solution trajectory strictly inside the admissible region and recovers the KKT-optimal policy as $\varepsilon_{\mathrm{bar}} \to 0$. This property provides a rigorous theoretical foundation for the learning-based and numerical procedures developed in the subsequent sections.

\section{Gradient-Based Algorithm for Policy Optimization}
\label{sec:BPTT_multiasset}

\subsection{Gradient-Based PMP Optimization with Barrier Regularization}
\label{sec:pgdpo} 

Building upon the barrier-regularized PMP framework developed in Section~\ref{sec:PMP_multiasset}, this section formulates a \emph{differential optimization procedure} that directly updates neural-network policies through gradient-based learning. We extend the unconstrained gradient-PMP approach of \citet{huh2025breaking} to fully \emph{constrained} portfolio–consumption problems. The key idea is to parameterize the portfolio and consumption policies using smooth neural networks and train them to approximately satisfy the Pontryagin first-order conditions while maintaining feasibility via the barrier-regularized Hamiltonian.

We refer to this method as \emph{Pontryagin-Guided Direct Policy Optimization} (PG-DPO). The approach follows the spirit of Direct Policy Optimization (DPO; e.g., \citealp{han2016deep}) but is fundamentally enhanced by an explicit correspondence between the backpropagation-through-time (BPTT) algorithm and the PMP adjoint system. As established in Theorem~\ref{thm:bptt_pmp_constrained}—which extends Theorem~3 of \citet{huh2025breaking} to the constrained setting considered here—the costates generated by BPTT constitute discrete-time representations of the PMP adjoint process. This rigorous connection justifies the ``Pontryagin-guided'' nomenclature and elevates the DPO framework from a heuristic approach to a theoretically grounded control method.

We encode inequality constraints in two complementary ways. First, feasibility is incorporated directly into the network architecture through smooth activation mappings at the output layer. This design preserves an end-to-end differentiable computation graph and removes the need to solve a separate optimization subproblem at each iteration. More specifically, all constraints are enforced via a log-barrier microproblem solved at each time–wealth pair \((t,x)\), ensuring that the stagewise updates remain on (an approximation of) the Pontryagin/KKT manifold. 

Second, when desired, we optionally apply a stagewise \emph{barrier projection} at deployment—referred to as \emph{Projected PG-DPO} (P-PGDPO)—which uses stabilized adjoints to align the learned controls with the barrier-regularized Hamiltonian. Crucially, the exponential–Euler rollout dynamics and the BPTT-based adjoint computations (described below) remain unchanged under this projection.

Let $(\boldsymbol{\pi}_{\theta}, C_{\phi})$ denote the parameterized portfolio and consumption policies with parameters $\theta$ and $\phi$, respectively, and let $\nu$ be a sampling distribution over a training domain 
$\mathcal D \subset [0, T)\times(0,\infty)$ of initial time–wealth pairs $(t_0, x_0)$. 
The training objective is defined as
\begin{equation}\label{eq:extended_objective_barrier}
J_{\mathrm e}(\theta,\phi)
=
\mathbb E_{(t_0,x_0)\sim\nu}\!
\left[
\mathbb E\!\left(
\int_{t_0}^{T}\!e^{-\rho(u-t_0)}U\!\bigl(C_\phi(u,X_u)\bigr)\,du
+ K\,e^{-\rho(T-t_0)}U(X_T)
\right)
\right],
\end{equation}
where feasibility of $(\boldsymbol{\pi}_{\theta}, C_{\phi})$ is enforced through smooth activation functions (and, if used, an interior-point projection), ensuring that $\Gamma_j(\boldsymbol{\pi}_{\theta}, C_{\phi}) > 0$ for all $j$. 

These activation maps transform unconstrained network outputs into feasible portfolio and consumption controls while remaining fully differentiable, thereby yielding an end-to-end differentiable objective $J_{\mathrm e}(\theta, \phi)$ that can be optimized via stochastic gradient ascent. Furthermore, the combination of smooth activation maps and the barrier term in the barrier-regularized Hamiltonian \eqref{eq:H_bar_again} guarantees KKT-consistent limiting behavior, as established in Theorem~\ref{thm:barrier_policy}.

We summarize below the activation functions used to guarantee strict feasibility and smooth gradients for all control variables:

\begin{itemize}
  \item \textbf{No borrowing and no short-sales.}  
  When both constraints are imposed simultaneously, we place the $n$ risky assets and the cash position ($i=0$) on the $(n+1)$-simplex using a softmax mapping:
  \begin{equation}\label{eq:softmax_simplex}
  \pi_i(t,x)
  =
  \frac{\exp\!\big(o_i(t,x)/\tau\big)}
  {\sum_{j=0}^{n}\exp\!\big(o_j(t,x)/\tau\big)}, 
  \qquad i=0,\dots,n.
  \end{equation}
  This parameterization ensures $\sum_{i=0}^n \pi_i = 1$ and $\pi_i \ge 0$ for all $i$. In particular, $\pi_0 \ge 0$ enforces the no-borrowing condition, while $\boldsymbol{\pi}_{1:n}\ge \mathbf{0}$ enforces the short-sale constraint.

 \item \textbf{Separated parameterization for borrowing vs.\ short-sales (optional).}  
To mirror the decomposition in Section~\ref{sec:dp_multiasset}—where borrowing is controlled by the cash weight and short-selling by risky weights—we parameterize the risky-asset allocation vector 
\[
\mathbf{w} = (w_1, \dots, w_n)^\top
\]
on the simplex and scale it by a \emph{budget share} $a \in (0,1)$. Specifically,
\begin{equation}\label{eq:softmax_risk_sig}
w_i(t,x) = \frac{\exp(\tilde o_i(t,x)/\tau)}{\sum_{j=1}^{n}\exp(\tilde o_j(t,x)/\tau)}, 
\quad
a(t,x) = \sigma(\tilde a(t,x)), 
\quad
\boldsymbol{\pi}_{1:n} = a\,\mathbf{w}, 
\quad 
\pi_0 = 1 - a.
\end{equation}
If borrowing is permitted but short-selling is not, one may replace $a = \sigma(\tilde a)$ with $a = \mathrm{softplus}(\tilde a)$, allowing $\pi_0 = 1 - a$ to become negative while still maintaining $\boldsymbol{\pi}_{1:n}\ge \mathbf{0}$.

  \item \textbf{Box-constrained consumption (including nonnegativity).}  
  For consumption subject to box constraints $C_{\min} \le C_t \le C_{\max}$, we map an unconstrained logit $v(t,x)\in\mathbb R$ to the admissible interval using a scaled hyperbolic tangent:
  \begin{equation}\label{eq:tanh_box}
  C(t,x) = \frac{C_{\max} + C_{\min}}{2} 
  + \frac{C_{\max} - C_{\min}}{2}\,\tanh v(t,x).
  \end{equation}
  The one-sided constraint $C_t \ge 0$ is obtained by setting $C_{\min}=0$. As a simple alternative, one may use a scaled softplus transformation, $C = \alpha\,\mathrm{softplus}(v)$ with $\alpha > 0$.
\end{itemize}

These activation functions are fully consistent with the interior-point formulation introduced in Section~\ref{sec:PMP_constrained}: they ensure that network outputs remain strictly feasible; when a log-barrier is used at deployment, the central path satisfies $\Gamma_j \ge c\,\varepsilon_{\mathrm{bar}}$ and converges to KKT as $\varepsilon_{\mathrm{bar}}\!\to\!0$.

Given the policy outputs $(\boldsymbol{\pi}_k, C_k)$ from the activation functions described above, the wealth dynamics evolve under the exponential–Euler discretization:
\begin{equation}\label{eq:exp_euler_wealth_act}
X_{k+1}
=
X_k \exp\!\Big(
\big[
\boldsymbol{\pi}_k^\top \widetilde{\boldsymbol{\mu}}
- \tfrac12\,\boldsymbol{\pi}_k^\top \widetilde{\boldsymbol{\Sigma}}\,\boldsymbol{\pi}_k
- C_k / X_k
\big] \Delta t
+ \boldsymbol{\pi}_k^\top \widetilde{\boldsymbol v}\,\Delta \mathbf{W}_k
\Big).
\end{equation}

We apply backpropagation through time (BPTT) with respect to the policy parameters $(\theta, \phi)$ through the wealth recursion \eqref{eq:exp_euler_wealth_act} and the activation mappings \eqref{eq:softmax_simplex}–\eqref{eq:tanh_box}. Automatic differentiation then provides discrete-time analogues of the Pontryagin costates,
\[
\lambda_k = \frac{\partial J_{\mathrm e}}{\partial X_k},
\qquad 
\mathbf Z_{k+1} = \frac{1}{\Delta t}\,\mathbb E\!\left[\lambda_{k+1}\,\Delta \mathbf W_k \,\big|\, \mathcal F_{t_k}\right],
\]
which correspond to the continuous-time adjoint dynamics 
$d\lambda_t = -\partial_X \mathcal H\,dt + \mathbf Z_t^\top d\mathbf W_t$.

The resulting stochastic gradient estimators are
\begin{align}
\nabla_\theta J_{\mathrm e}
&\approx
\mathbb E\!\left[
\sum_k
\Big\{
\lambda_k X_k \widetilde{\boldsymbol{\mu}}
+ X_k \big(\widetilde{\boldsymbol{v}}\,\mathbf Z_{k+1}\big)
\Big\}^\top
\frac{\partial \boldsymbol{\pi}_\theta(t_k, X_k)}{\partial \theta}\,\Delta t
\right], \label{eq:grad_theta_act} \\[0.6em]
\nabla_\phi J_{\mathrm e}
&\approx
\mathbb E\!\left[
\sum_k 
\Big(
-\,\lambda_k + e^{-\rho t_k} U'(C_k)
\Big)
\frac{\partial C_\phi(t_k, X_k)}{\partial \phi}\,\Delta t
\right]. \label{eq:grad_phi_act}
\end{align}

The BPTT recursion follows directly from the chain rule. Defining the discrete-time costate as
\[
\lambda_k := \frac{\partial J_e}{\partial X_k},
\]
we obtain the backward recursion
\begin{equation}\label{eq:bptt_chain_rule}
\lambda_k
= \frac{\partial R_k}{\partial X_k}
+ \mathbb{E}\!\left[\lambda_{k+1}\,\frac{\partial X_{k+1}}{\partial X_k}\,\Big|\,\mathcal F_{t_k}\right],
\qquad
R_k := e^{-\rho t_k} U(C_k)\,\Delta t .
\end{equation}

To compute $\frac{\partial X_{k+1}}{\partial X_k}$, it is convenient to rewrite the exponential–Euler scheme in an Euler–Maruyama form:
\[
X_{k+1} = X_k + b_k\,\Delta t + \sigma_k \cdot \Delta \mathbf{W}_k + r^X_k,
\]
where
\[
b_k := X_k\,\boldsymbol{\pi}_k^\top \widetilde{\boldsymbol{\mu}} - C_k,
\qquad
\sigma_k := X_k\,\boldsymbol{\pi}_k^\top \widetilde{\boldsymbol{v}},
\qquad
\mathbb E\!\big[\,|r^X_k|\,\big|\,\mathcal F_{t_k}\big] = O(\Delta t^{3/2}).
\]
It then follows that
\begin{equation}\label{eq:local_linearization:dx}
\frac{\partial X_{k+1}}{\partial X_k}
= 1 + \partial_x b_k\,\Delta t + \partial_x \sigma_k \cdot \Delta \mathbf{W}_k
  + O(\Delta t^{3/2}) .
\end{equation}
and the adjoint variable can be decomposed as
\begin{equation}\label{eq:local_linearization:mart}
\lambda_{k+1}
= \mathbb{E}\!\left[\lambda_{k+1}\,\big|\,\mathcal F_{t_k}\right]
  + \mathbf Z_{k+1}^{\!\top}\, \Delta \mathbf{W}_k .
\end{equation}

Here, the $O(\Delta t^{3/2})$ term represents the local truncation error associated with the Euler–Maruyama discretization. A precise error bound is provided in Theorem~\ref{thm:bptt_pmp_constrained}.

\begin{theorem}[BPTT--PMP correspondence with activation-induced feasible manifold]\label{thm:bptt_pmp_constrained}
Suppose Assumptions~\ref{ass:pmp_uncon} and~\ref{ass:bptt_pmp_reg} hold.
Let the portfolio/consumption controls be produced by smooth activation maps
$\Psi$ (e.g., \eqref{eq:softmax_simplex}--\eqref{eq:tanh_box}), and for each $(t,x)$ define the
\emph{activation-induced feasible fiber}
\[
\mathcal M_{t,x}\ :=\ \Psi(t,x;\mathbb R^p)
\ =\ \bigl\{(\boldsymbol{\pi},C)\in\mathbb R^{n+1}\times\mathbb R_+:\ 
\mathbf 1^\top\boldsymbol{\pi}=1,\ \Gamma(t,x;\boldsymbol{\pi},C)>\boldsymbol{0}\bigr\},
\]
assumed to be a $C^1$ embedded submanifold with tangent space
$T_{(\boldsymbol{\pi},C)}\mathcal M_{t,x}=\mathrm{Im}\,D_o\Psi(t,x;o)$ at $(\boldsymbol{\pi},C)=\Psi(t,x;o)$.
At time step $t_k$ and state $X_k$, the realized control $(\boldsymbol{\pi}_k,C_k)$ satisfies
$(\boldsymbol{\pi}_k,C_k)\in\mathcal M_{t_k,X_k}$ and the \emph{on-manifold first-order residual}
\[
\big\|\nabla_{\!\mathcal M}\mathcal H\!\left(t_k,X_k;\boldsymbol{\pi}_k,C_k,\lambda_{k+1},\mathbf Z_{k+1}\right)\big\|
\ \le\ \varepsilon_k,\qquad \sum_k \varepsilon_k\,\Delta t<\infty,
\]
where $\nabla_{\!\mathcal M}\mathcal H$ denotes the Riemannian (on-manifold) gradient of $\mathcal H$ on $\mathcal M_{t_k,X_k}$ (i.e., the restriction of $\nabla_{(\boldsymbol{\pi},C)}\mathcal H$ to feasible directions).
Define the discrete adjoints
\[
\lambda_k := \frac{\partial J_{\mathrm e}}{\partial X_k},\qquad
\mathbf Z_{k+1} := \frac{1}{\Delta t}\,\mathbb E\!\left[\lambda_{k+1}\,\Delta \mathbf W_k \,\big|\, \mathcal F_{t_k}\right],
\]
so that $\lambda_{k+1}=\mathbb E[\lambda_{k+1}\mid\mathcal F_{t_k}]+\mathbf Z_{k+1}^\top\Delta\mathbf W_k$.

\paragraph{(a) Discrete adjoint recursion.}
There exists a remainder $r_k$ with
$\mathbb E[r_k\mid\mathcal F_{t_k}]=O(\Delta t^{3/2}+\varepsilon_k\Delta t)$ such that
\[
\lambda_k-\lambda_{k+1}
=
\partial_x \mathcal H\!\left(t_k, X_k; \boldsymbol{\pi}_k, C_k, \lambda_{k+1}, \mathbf Z_{k+1}\right)\,\Delta t
\;-\; \mathbf Z_{k+1}^\top \Delta \mathbf W_k
\;+\; r_k .
\]

\paragraph{(b) $L^2$ convergence to the PMP adjoint.}
If $\Delta t\to 0$ and $\varepsilon_k\to 0$, the piecewise-constant interpolation of
$(\lambda_k,\mathbf Z_k)$ converges in $L^2$ to the unique solution $(\lambda_t,\mathbf Z_t)$ of
\[
d\lambda_t
=
-\partial_x \mathcal H\!\left(t, X_t; \boldsymbol{\pi}_t, C_t, \lambda_t, \mathbf Z_t\right)\,dt
+ \mathbf Z_t^\top d\mathbf W_t .
\]

\noindent\emph{Remark.} Throughout this theorem we work with the plain PMP Hamiltonian $\mathcal H$;
no log-barrier term appears in the BSDE driver. When used, the log-barrier enters only in the
deployment-time KKT projection (Section~\ref{sec:ppgdpo}).
\end{theorem}

\begin{proof}
See Appendix~\ref{app:proof_bptt_pmp}.
\end{proof}

\begin{remark}
Because feasibility is enforced directly by the activation mappings during training, no barrier terms appear in the BSDE driver. If desired, an additional log-barrier projection (P-PGDPO; see Section~\ref{sec:ppgdpo}) can be applied at deployment to refine the learned policy so that it lies exactly on the KKT manifold.
\end{remark}

Theorem~\ref{thm:bptt_pmp_constrained} establishes a rigorous theoretical bridge between the discrete learning dynamics induced by BPTT and the continuous-time adjoint system characterized by PMP. In essence, it shows that the gradients computed by standard automatic differentiation during neural policy training are not merely heuristic updates: in the limit of vanishing time steps, they coincide with the true continuous-time costate processes $(\lambda_t, \mathbf Z_t)$ solving the backward SDE
\[
d\lambda_t = -\partial_x \mathcal H(t, X_t; \boldsymbol{\pi}_t, C_t, \lambda_t, \mathbf Z_t)\,dt + \mathbf Z_t^\top d\mathbf W_t.
\]

This result generalizes the unconstrained BPTT--PMP correspondence of \citet{huh2025breaking} to the fully constrained setting considered here. The inclusion of activation-based feasibility mappings ensures that the control sequence $(\boldsymbol{\pi}_k, C_k)$ remains strictly within the admissible domain at all times, while the residual term $\varepsilon_k$ quantifies the local deviation from the exact Hamiltonian first-order condition at each step. The theorem implies that, as both the discretization step $\Delta t \to 0$ and the residuals $\varepsilon_k \to 0$, the discrete adjoints $(\lambda_k, \mathbf Z_k)$ converge in $L^2$ to the continuous-time adjoint pair satisfying the PMP conditions.

This correspondence provides a principled justification for training neural policies via gradient backpropagation rather than solving high-dimensional FBSDEs explicitly. It guarantees that the gradient signals propagated through the computational graph are consistent with the underlying optimal control theory of PMP. The BPTT recursion reproduces the discrete analogue of the costate dynamics governed by $\partial_x \mathcal H$, and the estimated gradient directions \eqref{eq:grad_theta_act}–\eqref{eq:grad_phi_act} coincide asymptotically with those derived from the PMP first-order conditions. Consequently, each gradient-based update moves the policy parameters in the same direction as an infinitesimal adjustment along the PMP manifold.

In practical training, these results guarantee that PG-DPO behaves as a faithful discrete approximation of the continuous-time PMP optimizer. When the time step $\Delta t$ is sufficiently small and the policy networks are expressive enough, the discrepancy between BPTT-based learning and the exact PMP control law is on the order of $O(\Delta t^{3/2}) + O(\varepsilon_k \Delta t)$. Consequently, the learning dynamics produced by PG-DPO track the true continuous-time adjoint system with vanishing error as the rollout resolution increases. This theoretical property sharply distinguishes PG-DPO from purely heuristic reinforcement-learning updates: it is not an empirical surrogate but a principled, PMP-consistent discretization of continuous-time optimal control.

Moreover, the combination of Theorem~\ref{thm:bptt_pmp_constrained} and Theorem~\ref{thm:barrier_policy} (from Section~\ref{sec:PMP_multiasset}) completes the theoretical foundation of our framework. In particular, the barrier-regularized PMP ensures feasibility and convergence to the KKT solution, while the BPTT--PMP correspondence guarantees that the gradient-based training algorithm faithfully reproduces the continuous-time costate dynamics necessary for Pontryagin-optimal control. Together, these results establish PG-DPO as a fully differentiable, theoretically grounded method for constrained continuous-time portfolio optimization.

Finally, while the barrier-regularized PG-DPO ensures differentiable feasibility and Pontryagin-consistent gradients during training, it can be further refined by explicitly projecting the learned controls onto the barrier-regularized Hamiltonian manifold. Section~\ref{sec:ppgdpo} introduces this projection-based variant, termed \emph{Projected PG-DPO (P-PGDPO)}, which improves numerical stability and alignment with the KKT stationary conditions in practice.

\subsection{Projected PG-DPO (P-PGDPO) via Log-Barrier Projection}
\label{sec:ppgdpo}

Projected PG-DPO (P-PGDPO) refines the baseline PG-DPO by decoupling learning and deployment. The method first learns accurate adjoints (costates) for the continuous-time PMP system and then, at deployment, recovers the control at each time–wealth pair by solving a small log-barrier–regularized stagewise maximization. This separation targets the main statistical and computational bottleneck in high dimensions: emulating the \emph{stagewise} Pontryagin maximizer with a single policy network is demanding because the control must depend on $(t,x)$, model coefficients, and adjoints while also satisfying pointwise feasibility. By asking the network only to learn smooth, expressive adjoints and leaving the constrained maximization to a local solver, P-PGDPO exploits the favorable local second-order geometry of the stagewise problem. The log barrier supplies diagonal curvature and keeps the iterate strictly interior, which stabilizes Newton steps near active-set transitions (see Section~\ref{sec:PMP_constrained}). As $\varepsilon_{\mathrm{bar}}\downarrow 0$, the barrier solution reduces to the KKT projection; in this sense, the barrier is a smooth regularization of the same stagewise optimizer with the same Pontryagin-consistent limit.

The procedure has two stages.

\medskip\noindent
\textbf{Stage 1 (warm-up and autodiff costate estimation).} Run the baseline PG-DPO of Section~\ref{sec:pgdpo} for $K_0$ iterations with smooth feasibility-preserving activations. Automatic differentiation yields
\[
\lambda_k=\frac{\partial J_e}{\partial X_k},\qquad
\partial_x\lambda_k=\frac{\partial^2 J_e}{\partial X_k^2}.
\]
After this warm-up, freeze $(\theta,\phi)$ and form Monte Carlo averages to stabilize the adjoint estimates:
\[
\widehat{\lambda}(t,x)\approx \frac{1}{M_{\!MC}}\sum_{j=1}^{M_{\!MC}}\lambda^{(j)}(t,x),
\qquad
\partial_x\widehat{\lambda}(t,x)\ \text{analogously}.
\]
For deployment we use the Markov identity from Section~\ref{sec:PMP_unconstrained},
\[
\widehat{\mathbf Z}(t,x;\boldsymbol\pi)
=\big(\partial_x\widehat{\lambda}\big)(t,x)\cdot x\,\widetilde{\boldsymbol v}^{\!\top}\boldsymbol\pi,
\]
so there is no need to regress a separate $\widehat{\mathbf Z}$ network. No barrier terms appear in the BSDE driver during training; feasibility is already enforced by the activations.

\medskip\noindent
\textbf{Stage 2 (log-barrier projection).} Given $(\widehat{\lambda},\partial_x\widehat{\lambda})$, compute the deployment-time control at any state $(t,x)$ by solving the barrier-regularized stagewise maximization
\begin{equation}\label{eq:proj_controls}
\begin{aligned}
\mathbf u^{\mathrm{proj}}(t,x)
&=\operatorname*{arg\,max}_{\boldsymbol\pi,\,C,\,\eta}\ 
\widetilde{\mathcal H}_{\mathrm{bar}}\big(t,x;\boldsymbol\pi,C,\eta;\widehat{\lambda},\widehat{\mathbf Z}\big)\\[0.2em]
&\text{s.t.}\ \Gamma_j\big(t,x;\boldsymbol\pi,C\big)>0,\quad j=1,\dots,m,
\end{aligned}
\end{equation}
with
\[
\widetilde{\mathcal H}_{\mathrm{bar}}
=\mathcal H\!\big(t,x;\boldsymbol\pi,C,\widehat{\lambda},\widehat{\mathbf Z}\big)
+\eta\,(1-\mathbf 1^\top\boldsymbol\pi)
+\varepsilon_{\mathrm{bar}}\sum_{j=1}^m \ln\Gamma_j\!\big(t,x;\boldsymbol\pi,C\big),
\qquad \varepsilon_{\mathrm{bar}}>0,\ \eta\in\mathbb R.
\]
We solve \eqref{eq:proj_controls} by a few Newton–CG iterations on the tangent space $\{\mathbf 1^\top d=0\}$ (or, equivalently, via a $1\times 1$ Schur complement in $\eta$), combined with a fraction-to-the-boundary line search and an Armijo sufficient-increase test as in Section~\ref{sec:PMP_constrained}. The barrier injects diagonal curvature proportional to $\varepsilon_{\mathrm{bar}}/\Gamma_j^2$ and ensures strict interiority $\Gamma_j\ge c\,\varepsilon_{\mathrm{bar}}$, which prevents numerical stalls around active-set switches. It is neither necessary nor desirable to oversolve this microproblem: the KKT residual tolerance $\delta_{\mathrm{proj}}$ should be set on the same or smaller scale as the adjoint-estimation error. With this choice, Stage~2 contributes negligible error relative to Stage~1 while preserving excellent conditioning.

The following result formalizes how the total policy error decomposes into controllable pieces and clarifies the role of the barrier as a central-path bias.

\begin{theorem}[Policy-gap bound for P-PGDPO under log-barrier projection (revised)]
\label{thm:ppgdpo_gap_barrier}
Suppose Assumptions~\ref{ass:Gamma}, \ref{ass:barrier_kkt}, and~\ref{ass:barrier_setup_appendix} hold.
Let $\varepsilon_{\mathrm{foc}}$ be the warm-up Pontryagin FOC residual (in an $L^{q,p}$ norm) and set $\delta_{\mathrm{bptt}}\equiv \kappa_1\Delta t+\kappa_2/\sqrt{M}$. If $\delta_{\mathrm{proj}}$ is the Stage~2 residual tolerance and $\varepsilon_{\mathrm{bar}}>0$ is the barrier parameter, then there exist constants
$C_{\mathrm{proj}},C_{H,1},C_{H,2},C_{H,3}>0$ (uniform on the deployment domain) such that
\[
\|\boldsymbol\pi^{\mathrm{proj}}-\boldsymbol\pi^*\|_{L^{q,p}}
\ \le\
C_{\mathrm{proj}}\big(\varepsilon_{\mathrm{foc}}+\delta_{\mathrm{bptt}}+\delta_{\mathrm{proj}}+\varepsilon_{\mathrm{bar}}\big),
\]
and, for each time--state pair, the instantaneous Hamiltonian gap satisfies the \emph{mixed-order} bound
\begin{align*}
0\ \le\ \mathcal H(\boldsymbol\pi^*,\cdot)-\mathcal H(\boldsymbol\pi^{\mathrm{proj}},\cdot)
&\ \le\ C_{H,1}\,\varepsilon_{\mathrm{bar}}
\;+\;C_{H,2}\big(\varepsilon_{\mathrm{foc}}+\delta_{\mathrm{bptt}}+\delta_{\mathrm{proj}}\big)\\
&\qquad\qquad
\;+\;C_{H,3}\big(\varepsilon_{\mathrm{foc}}^{\,2}+\delta_{\mathrm{bptt}}^{\,2}+\delta_{\mathrm{proj}}^{\,2}\big).
\end{align*}
In particular, if no inequality constraint is active at the KKT solution, the $O(\varepsilon_{\mathrm{bar}})$ term vanishes and the bound tightens to
\[
\mathcal H(\boldsymbol\pi^*,\cdot)-\mathcal H(\boldsymbol\pi^{\mathrm{proj}},\cdot)
\ \le\ \widetilde C_{H,1}\,\varepsilon_{\mathrm{bar}}^{2}
\;+\;\widetilde C_{H,2}\big(\varepsilon_{\mathrm{foc}}^{\,2}+\delta_{\mathrm{bptt}}^{\,2}+\delta_{\mathrm{proj}}^{\,2}\big).
\]
\end{theorem}

\begin{proof}
See Appendix~\ref{app:proof_ppgdpo}.
\end{proof}

\begin{corollary}[Consistency and rate separation (revised)]
\label{cor:ppgdpo_consistency}
If $\varepsilon_{\mathrm{foc}}\to0$, $\Delta t\to0$, $M\to\infty$ (so $\delta_{\mathrm{bptt}}\to0$), $\delta_{\mathrm{proj}}\to0$, and $\varepsilon_{\mathrm{bar}}\to0$, then $\boldsymbol\pi^{\mathrm{proj}}\to \boldsymbol\pi^*$ in $L^{q,p}$ and, uniformly on the deployment domain,
\[
\mathcal H(\boldsymbol\pi^*,\cdot)-\mathcal H(\boldsymbol\pi^{\mathrm{proj}},\cdot)
\ \le\ 
C_{H,1}\,\varepsilon_{\mathrm{bar}}
\;+\;C_{H,2}\big(\varepsilon_{\mathrm{foc}}+\delta_{\mathrm{bptt}}+\delta_{\mathrm{proj}}\big)
\;+\;C_{H,3}\big(\varepsilon_{\mathrm{foc}}^{\,2}+\delta_{\mathrm{bptt}}^{\,2}+\delta_{\mathrm{proj}}^{\,2}\big)
\ \xrightarrow[]{} 0.
\]
In particular, if the KKT solution is strictly interior (no active inequality constraints), then the linear term vanishes and the Hamiltonian gap admits the purely quadratic rate
\[
\mathcal H(\boldsymbol\pi^*,\cdot)-\mathcal H(\boldsymbol\pi^{\mathrm{proj}},\cdot)
\ \le\ 
\widetilde C_{H,1}\,\varepsilon_{\mathrm{bar}}^{2}
\;+\;\widetilde C_{H,2}\big(\varepsilon_{\mathrm{foc}}^{\,2}+\delta_{\mathrm{bptt}}^{\,2}+\delta_{\mathrm{proj}}^{\,2}\big).
\]
For a fixed computational budget, it is effective to choose
$\delta_{\mathrm{proj}}\ll \varepsilon_{\mathrm{foc}}+\delta_{\mathrm{bptt}}$ and to schedule $\varepsilon_{\mathrm{bar}}$ so that
$C_{H,1}\varepsilon_{\mathrm{bar}}\lesssim C_{H,2}\big(\varepsilon_{\mathrm{foc}}+\delta_{\mathrm{bptt}}\big)$,
decreasing $\varepsilon_{\mathrm{bar}}$ until numerical conditioning begins to deteriorate.
\end{corollary}

Two practical remarks place the method in context. First, the deployment-time solve is tiny—only $(n\!+\!2)$ variables—so each barrier step is extremely fast, and the overall cost per time step scales linearly in the number of assets $n$. Second, a simple geometric schedule for $\varepsilon_{\mathrm{bar}}$ with a small lower floor works well: the diagonal barrier curvature stabilizes Newton steps around active-set switches while the central-path bias vanishes as $\varepsilon_{\mathrm{bar}}\downarrow 0$. Altogether, the barrier view can be read as a numerically robust regularization of KKT projection with the same Pontryagin-consistent limit and with explicit, tunable error decomposition via Theorem~\ref{thm:ppgdpo_gap_barrier} and Theorem~\ref{thm:barrier_policy}.

\section{Constrained Experiments: Short-Sale Ban and Consumption Cap}\label{sec:num_test}

Our experiments focus on two canonical constrained settings designed to evaluate the effectiveness of Pontryagin-guided projection in enforcing feasibility and recovering accurate controls. The aim is not to provide an exhaustive empirical benchmark across all possible solvers, but rather to demonstrate that the proposed policy-centric framework produces PMP/KKT-consistent solutions precisely in scenarios where value-based approaches tend to fail.

We examine two practically important classes of constraints.  
First, we impose a \emph{short-sale ban} on risky-asset weights, requiring all risky positions to remain non-negative while maintaining full investment by adjusting the residual cash position.  
Second, we enforce a \emph{consumption cap} specified as a fixed fraction of wealth, which introduces a hard upper bound on spending at each time–state pair.  
For each constraint, we compare two algorithmic variants:  
(i) the baseline PG-DPO, which enforces feasibility only through differentiable activation functions, and  
(ii) the P-PGDPO variant, which explicitly projects the learned controls onto the Pontryagin/KKT manifold using a barrier-regularized Hamiltonian.

Throughout all tables and figures, $\boldsymbol{\pi}_{1:n}$ denotes the vector of portfolio weights in risky assets, and full investment is enforced by setting the residual cash allocation as $\pi_0 = 1 - \mathbf{1}^\top \boldsymbol{\pi}_{1:n}$.  
The \emph{reference policy}—used as the ground truth for evaluating approximation error—is the analytical or numerically exact constrained Pontryagin/KKT solution computed directly from the true model parameters and adjoint processes.

%Specifically, for the short-sale benchmark, the reference policy follows the piecewise closed-form KKT solution for convex portfolio constraints \citet{cvitanic1993hedging} and \citet{XuShreve1992a,XuShreve1992b, karatzas1998methods}. For the consumption-cap benchmark $0 \le C_t \le \bar m X_t$, the reference is the standard ``clipping'' rule, $C_t^*=\min\{\max\{(U')^{-1}(e^{\rho t}\lambda_t),\,0\},\,\bar m X_t\}$, which arises directly from the PMP first-order conditions for box-constrained controls \citet{fleming2006controlled} and \citet{pham2009continuous}.

\subsection{Experimental Scope}\label{sec:benchmark_scope}

Before presenting the numerical results, it is useful to clarify the scope and motivation of our experimental design. Classical dynamic programming (DP) methods, while theoretically well understood, become computationally infeasible beyond very low dimensions due to the curse of dimensionality. Similarly, value-based neural PDE solvers—such as deep BSDE approaches \citep{han2018solving,e2017deep} and physics-informed neural networks (PINNs) \citep{raissi2019physics}—remain fundamentally tied to the value-function representation. As discussed in Section~\ref{sec:value_limitations}, once inequality constraints are introduced, the HJB equation transforms into a variational inequality or free-boundary PDE, and the corresponding BSDE becomes a reflected BSDE. These transitions introduce severe analytical and numerical difficulties.

A growing body of literature documents these limitations. For example, \citet{andersson2023convergence} show that deep BSDE methods can fail to converge even in simple linear–quadratic control settings. \citet{iftakher2025physics} and \citet{krishnapriyan2021characterizing} report persistent constraint violations and ill-conditioning in PINNs under inequality or free-boundary conditions, while practical studies such as \citet{tsang2020deep} often simplify the problem formulation or hard-wire constraints into the architecture to bypass these issues.

Against this backdrop, the goal of our experiments is not to benchmark against every possible numerical solver, but rather to demonstrate that the proposed policy-centric, Pontryagin-guided framework is both theoretically principled and practically effective precisely in the regimes where value-based approaches break down. The experiments in Sections~\ref{sec:ssban} and \ref{sec:cons_cap} are therefore deliberately minimalistic. They are designed to isolate the effects of key constraints—namely, short-sale bans and consumption caps—and to show that Pontryagin-guided projection produces accurate, feasible controls in scenarios where value-based baselines face intrinsic difficulties.

A distinctive strength of our approach lies in its scalability to high-dimensional, constrained portfolio problems that are far beyond the reach of classical PDE- or BSDE-based solvers. Because PG-DPO and P-PGDPO operate directly in the \emph{policy space}—rather than discretizing the value function—the computational complexity scales \emph{linearly} with the number of assets, rather than exponentially. This enables, for the first time, the accurate and feasible learning of constrained optimal policies in environments with up to 100 risky assets.\footnote{In additional stress tests, we confirm that PG-DPO can be applied to environments with as many as 1000 risky assets, demonstrating the scalability of the approach.} 

The experiments that follow are designed to highlight this property. We show that Pontryagin-guided methods remain numerically stable, strictly feasible, and recover KKT-consistent controls even in large-scale multi-asset settings where conventional HJB, PINN, or deep-BSDE formulations either diverge or become computationally intractable.

\subsection{Short-Sale Constraint (No Shorting)}\label{sec:ssban}

\noindent
We begin with the short-sale ban, one of the most prevalent and practically important constraints in portfolio management. The investor allocates wealth across $n$ risky assets and one risk-free asset subject to the full-investment condition
\[
\mathbf 1^\top \boldsymbol{\pi} = 1,
\]
together with the inequality constraints
\[
\pi_i \ge 0, \qquad i = 1, \dots, n,
\]
which prohibit short positions in any risky asset. The residual cash allocation is determined endogenously as $\pi_0 = 1 - \mathbf 1^\top \boldsymbol{\pi}_{1:n}$. 

This experiment is designed to assess how accurately the proposed Pontryagin-guided algorithms recover feasible, KKT-consistent controls under these inequality constraints, particularly in high-dimensional settings where classical approaches become unreliable. Our focus is on evaluating both feasibility (strict satisfaction of the non-negativity and full-investment conditions) and optimality (closeness to the true Pontryagin solution) as the number of risky assets increases.

\begin{table}[t!]
\centering
\caption{No short-selling (non-negative risky weights): RMSE of risky-weight vector $\boldsymbol{\pi}_{1:n}$ versus the PMP/KKT reference (lower is better).}
\label{tab:ss_rmse}
\begin{tabular}{rcc}
\toprule
\(\mathbf{n}\) & \textbf{PG-DPO} & \textbf{P-PGDPO} \\
\midrule
2 & 0.017035 & 0.002509 \\
10 & 0.079714 & 0.001251 \\
100 & 0.011093 & 0.003015 \\
\bottomrule
\end{tabular}
\end{table}

The benchmark environment is based on the constant-coefficient multi-asset Merton model introduced in Section~\ref{sec:dp_multiasset}. The investor’s preferences are represented by a constant-relative-risk-aversion (CRRA) utility function,
\[
U(C) = \frac{C^{1-\gamma}}{1-\gamma}, \qquad \gamma \neq 1,
\]
with fixed parameters: risk-free rate $r = 0.03$ (per annum), subjective discount rate $\rho = 0.10$, and risk-aversion coefficient $\gamma = 2.0$. Training trajectories are generated under an Euler--Maruyama discretization with time step $\Delta t = 0.075$ and investment horizon $T = 1.5$.

We consider both moderate-dimensional environments ($n=2,10$) and a large-scale setting ($n=100$). 
In our implementation, policy networks are plain MLPs with two hidden layers of width $200$ and LeakyReLU activations. 
Feasibility is enforced through output maps tailored to each constraint: for the no-short experiment we apply a softplus on the risky-asset weights to ensure nonnegativity.

For evaluation, the \emph{reference policy} is taken to be the analytical KKT-optimal portfolio derived from the constrained Merton problem, as characterized by \citet{XuShreve1992a, XuShreve1992b}. This allows us to directly measure the policy error  relative to the true optimal control.

Table~\ref{tab:ss_rmse} reports the root mean square error (RMSE) of the risky-asset weight vector $\boldsymbol{\pi}_{1:n}$ for portfolio dimensions $n \in \{2,10,100\}$. The RMSE, defined as the square root of the mean squared deviation between the learned policy and the reference KKT-optimal policy, provides a direct measure of the typical magnitude of policy error. 

Across all dimensions, the projection-based P-PGDPO variant consistently outperforms the baseline PG-DPO. Relative to the activation-only approach, P-PGDPO reduces the $\boldsymbol{\pi}_{1:n}$-RMSE by approximately $85\%$ when $n=2$, $98\%$ when $n=10$, and $73\%$ when $n=100$. These improvements remain substantial even in the high-dimensional setting, underscoring the stability and effectiveness of the projection mechanism.

The advantage of P-PGDPO becomes more pronounced as dimensionality increases. This is consistent with the fact that activation-only feasibility leaves behind an $\varepsilon$-stationarity bias in the Hamiltonian first-order condition—a bias that becomes increasingly difficult to eliminate as the number of assets grows. By contrast, the barrier-based projection step explicitly recovers the stagewise Pontryagin maximizer on the feasible manifold and exploits the diagonal curvature induced by the barrier term, leading to sharper convergence and more accurate policy recovery in large-scale environments.

\begin{figure}[t!]
\centering
\begin{subfigure}{.32\textwidth}
\centering
\includegraphics[width=\linewidth]{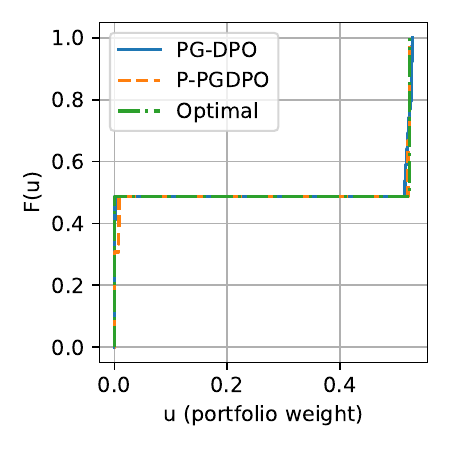}
\caption{\(n=2\)}
\end{subfigure}\hfill
\begin{subfigure}{.32\textwidth}
\centering
\includegraphics[width=\linewidth]{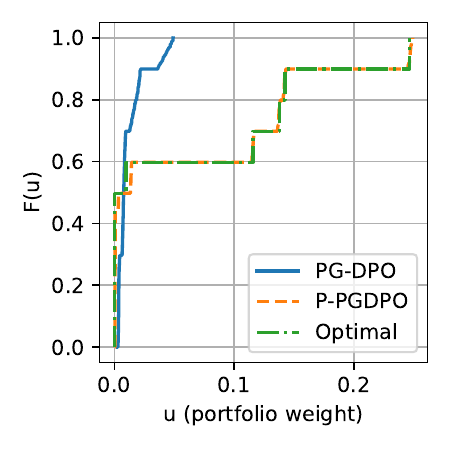}
\caption{\(n=10\)}
\end{subfigure}\hfill
\begin{subfigure}{.32\textwidth}
\centering
\includegraphics[width=\linewidth]{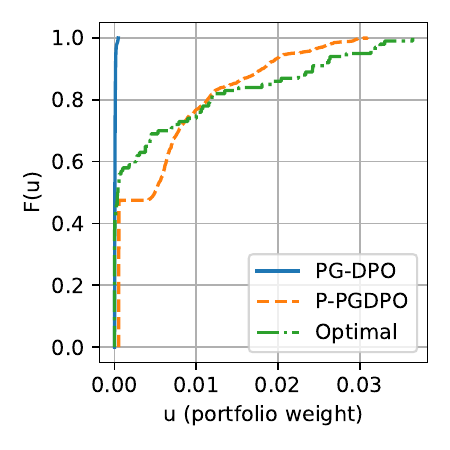}
\caption{\(n=100\)}
\end{subfigure}
\caption{No short-selling: ECDFs of pooled risky weights $\boldsymbol{\pi}_{1:n}$ (over time and assets) for PG-DPO, P-PGDPO (manifold projection), and the reference policy. Closer overlap with the reference ECDF implies more accurate recovery of the target $\boldsymbol{\pi}_{1:n}$-control.}
\label{fig:ss_ecdf}
\end{figure}

Figure~\ref{fig:ss_ecdf} compares the empirical cumulative distribution functions (ECDFs) of the pooled risky-asset weights across all time steps, assets, and Monte Carlo paths. The ECDF is a nonparametric estimate of the cumulative distribution function: for a set of observations, its value at $x$ is the fraction of samples less than or equal to $x$. In our setting, the observations are the risky-asset weights $\pi_{i,t}$, pooled over all assets $i=1,\dots,n$, all time steps $t$, and all Monte Carlo trajectories. This pooled distribution provides a comprehensive view of how portfolio mass is allocated over time and across assets—effectively indicating how often and how heavily each risky position is held. 

Closer alignment of the learned ECDF to the analytical benchmark—given either by the closed-form solution $\boldsymbol{\pi}_{\mathrm{cf},1:n}$ in the unconstrained limit or by its constrained Pontryagin/KKT projection $\boldsymbol{\pi}_{\mathrm{pp},1:n}$—indicates that the policy successfully recovers the correct $\boldsymbol{\pi}_{1:n}$ control. As shown in Figure~\ref{fig:ss_ecdf}, P-PGDPO tracks the reference distribution remarkably well across the entire support, including the tails, whereas the baseline PG-DPO exhibits systematic deviations, accumulating excessive mass away from the optimal allocation as the dimensionality $n$ increases. This distributional consistency complements the RMSE results by demonstrating that the projection-based approach not only reduces average errors but also recovers the entire empirical distribution of risky weights with high fidelity.

Overall, the short-sale experiments demonstrate that the barrier-based projection in P-PGDPO reliably recovers the stagewise Pontryagin maximizer on the feasible manifold. By construction, the projection step eliminates constraint violations, drives the Hamiltonian residuals toward zero, and preserves linear computational scaling with respect to the number of assets $n$. In contrast, the baseline PG-DPO exhibits mild feasibility drift as dimensionality increases, reflecting the accumulation of stationarity error when feasibility is enforced solely through smooth activations. Crucially, even in high-dimensional environments with $n=100$, P-PGDPO achieves accuracy nearly indistinguishable from the analytical reference solution, underscoring the scalability and robustness of the policy-centric approach. Because the underlying computational graph scales linearly in $n$, the same architecture can, in principle, be extended to portfolios with several hundred—or even thousands—of assets without fundamental modification.

\subsection{Consumption Cap (as a Fraction of Wealth)}\label{sec:cons_cap}

We next examine a pointwise consumption constraint of the form
\[
0 \ \le\ C_t \ \le\ \bar m\,X_t,
\]
with a fixed cap $\bar m\in(0,1)$. This condition restricts consumption to a fixed fraction of current wealth, capturing liquidity or prudence constraints frequently imposed in practice. Feasibility is ensured either by a smooth activation that maps the consumption logit into the admissible interval $[0,\bar m X_t]$ or by solving a barrier-regularized microproblem that projects the stagewise control onto the Pontryagin/KKT manifold. Although the consumption control $C_t$ is one-dimensional—so the absolute performance gains are less dramatic than in the high-dimensional portfolio block—it remains tightly coupled with the portfolio choice. In particular, sharpening the $\boldsymbol{\pi}_{1:n}$ control through manifold projection stabilizes the stagewise Hamiltonian near the active boundary and leads to smoother, more accurate consumption trajectories. 

The theoretical reference for this benchmark is the analytical consumption policy implied by the Pontryagin/KKT first–order conditions under box constraints:
\[
C_t^*=\min\!\Bigl\{\max\!\Bigl\{(U')^{-1}\!\bigl(e^{\rho t}\lambda_t\bigr),\,0\Bigr\},\,\bar m X_t\Bigr\},
\]
as derived in \citet{fleming2006controlled} and \citet{pham2009continuous}. Throughout this subsection we keep the market/utility parameters, time horizon and discretization, and the drift–volatility specification identical to those used in the short-sale experiment in Section~\ref{sec:ssban}; only the control parameterization changes to include consumption. This analytical $C_t^*$ serves as the ground-truth target against which we evaluate the learned consumption policies.

\begin{table}[t!]
\centering
\caption{Consumption cap $0\le C_t\le \bar m X_t$: RMSE (lower is better) for risky weights $\boldsymbol{\pi}_{1:n}$ (left) and consumption $C$ (right).}
\label{tab:cap_rmse_both}
\begin{subtable}[t]{.48\textwidth}
\centering
\caption{$\boldsymbol{\pi}_{1:n}$-RMSE}
\label{tab:cap_u_rmse}
\begin{tabular}{rcc}
\toprule
\(\mathbf{n}\) & \textbf{PG-DPO} & \textbf{P-PGDPO} \\
\midrule
2 & 0.033858 & 0.000321 \\
10 & 0.055776 & 0.000313 \\
100 & 0.038807 & 0.000518 \\
\bottomrule
\end{tabular}
\end{subtable}\hfill
\begin{subtable}[t]{.48\textwidth}
\centering
\caption{$C$-RMSE}
\label{tab:cap_c_rmse}
\begin{tabular}{rcc}
\toprule
\(\mathbf{n}\) & \textbf{PG-DPO} & \textbf{P-PGDPO} \\
\midrule
2 & 0.234102 & 0.188612 \\
10 & 0.225106 & 0.188153 \\
100 & 0.223304 & 0.188417 \\
\bottomrule
\end{tabular}
\end{subtable}
\end{table}

Table~\ref{tab:cap_rmse_both} reports the RMSE for both the risky-asset weights and consumption across dimensions $n\in\{2,10,100\}$. The manifold-projection variant (P-PGDPO) reduces the $\boldsymbol{\pi}_{1:n}$-RMSE by nearly two orders of magnitude at all dimensions, while improving the $C_t$-RMSE by approximately $15\text{--}19\%$ relative to the baseline PG-DPO. The disparity in relative improvement is expected: the consumption decision is scalar and typically operates near its upper boundary, leaving less numerical room for large accuracy gains than the high-dimensional portfolio block. Nevertheless, because the first-order condition for $C_t$ is coupled to $\boldsymbol{\pi}$ through the Hamiltonian, a more accurate recovery of the stagewise Pontryagin maximizer in the $\boldsymbol{\pi}_{1:n}$-block reduces both the bias and variance of $C_t$. This interdependence highlights a key advantage of the policy-centric formulation: improving the portfolio control indirectly enhances the quality of the consumption decision, even when the latter is subject to tight box constraints.

\begin{figure}[t!]
\centering
\begin{subfigure}[t]{0.8\textwidth}
\centering
\includegraphics[width=\linewidth]{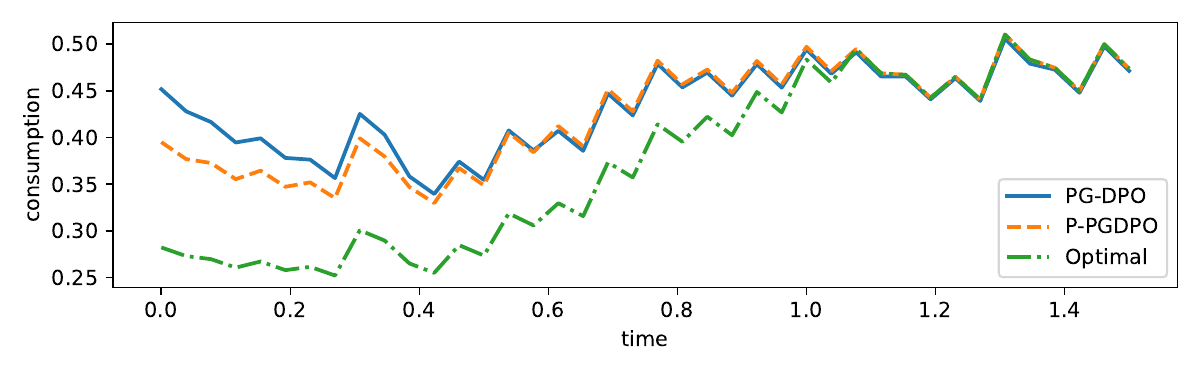}
\caption{\(n=2\)}
\end{subfigure}

\begin{subfigure}[t]{0.8\textwidth}
\centering
\includegraphics[width=\linewidth]{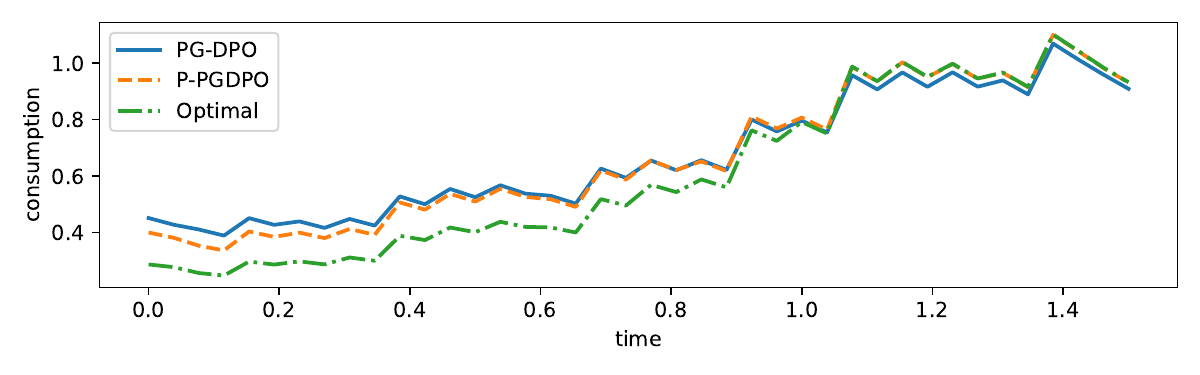}
\caption{\(n=10\)}
\end{subfigure}

\begin{subfigure}[t]{0.8\textwidth}
\centering
\includegraphics[width=\linewidth]{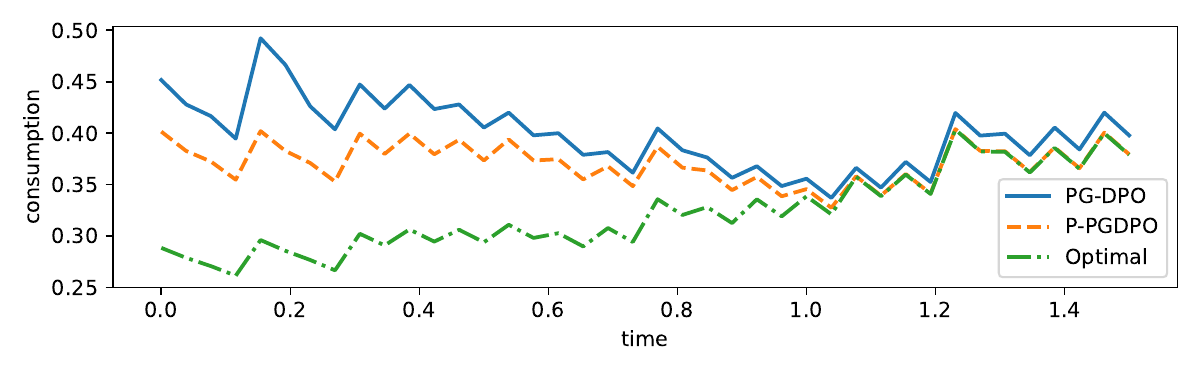}
\caption{\(n=100\)}
\end{subfigure}
\caption{Consumption paths under the cap $C_t\le \bar m X_t$ for PG-DPO, P-PGDPO (manifold projection), and the reference policy. P-PGDPO is uniformly closer to the reference and reduces boundary over/undershoot, but it is \emph{not} exact: residual deviations remain near binding times and turning points.}
\label{fig:cap_cons_paths}
\end{figure}

Figure~\ref{fig:cap_cons_paths} illustrates representative consumption trajectories for a single Monte Carlo path under the cap constraint. Across all dimensions, the manifold-projection variant (P-PGDPO) tracks the reference policy much more closely and exhibits markedly smaller over- and undershoots near the upper boundary. By contrast, the baseline PG-DPO displays more pronounced boundary-layer deviations, particularly at early times. This behavior arises because the terminal condition anchors the backward-propagated costate signals used in policy optimization: as time progresses backward from $T$, the cumulative estimation error in the adjoints grows, making the learning problem increasingly challenging. Near maturity, both methods benefit from stronger and more accurate terminal signals, resulting in a reduced gap between the two algorithms.

Despite these improvements, the learned consumption paths do not perfectly coincide with the analytical reference. Small residual deviations persist, especially when the cap is binding or near turning points in the optimal consumption schedule. These residuals reflect the inherently one-dimensional nature of $C_t$, the non-smooth kink introduced by the box constraint, and the accumulated adjoint-estimation error discussed above. Nevertheless, the reduced boundary oscillations and closer adherence to the reference trajectory highlight the stabilizing effect of the Pontryagin-guided projection in constrained consumption problems.

In summary, although the consumption control is inherently one-dimensional and its relative performance gains are smaller than those observed in the high-dimensional portfolio block, the barrier-projected formulation nonetheless delivers substantial benefits by solving the joint $(\boldsymbol{\pi}_t, C_t)$ maximization more accurately. The manifold-projection variant consistently enhances numerical stability near the binding boundary and produces smoother consumption trajectories over time. Across all tested dimensions, P-PGDPO achieves near-reference accuracy for both risky weights and consumption, demonstrating that the proposed policy-centric framework remains robust and effective even when portfolio and consumption decisions are tightly coupled through nonlinear Hamiltonian constraints.

\section{Conclusion}\label{sec:conclusion}

This paper has introduced a scalable, Pontryagin-guided, policy-centric framework for continuous-time, high-dimensional portfolio optimization under realistic inequality constraints. By embedding the first-order optimality structure of Pontryagin’s Maximum Principle (PMP) directly into a differentiable learning architecture, we developed the PG-DPO algorithm and its barrier-projected variant, P-PGDPO, which jointly guarantee feasibility, numerical stability, and theoretical convergence to the KKT solution. The barrier-regularized Hamiltonian formulation provides a smooth interior-point representation of constrained optimal controls, while Theorem~\ref{thm:barrier_policy} establishes a formal correspondence between discrete BPTT adjoints and the continuous-time PMP costates. Together, these results close a longstanding gap between analytical stochastic control theory and modern neural policy optimization.

Empirically, the proposed framework accurately recovers constrained PMP/KKT solutions in two canonical benchmark settings: a short-sale ban on risky assets and a wealth-proportional consumption cap. Across all dimensions, the manifold-projection variant P-PGDPO enforces strict feasibility and first-order optimality, reducing policy error by up to two orders of magnitude relative to the baseline activation-only PG-DPO. Crucially, the computational complexity of both algorithms scales linearly with the number of assets, enabling accurate and feasible training in large-scale environments with 100 or even 1000 assets—a regime where classical DP, HJB, or deep-BSDE methods become computationally intractable.

A distinctive advantage of our approach lies in its constructive nature even when classical solutions are unavailable. When closed-form constrained policies are unknown—or when classical optimality fails under pathological caps or interacting constraints—the barrier central path yields well-defined, feasible surrogate controls with monotone Hamiltonian ascent and explicitly quantifiable error bounds. These properties make the proposed framework a practical and theoretically grounded tool precisely in the regimes where DP grids are infeasible, free-boundary formulations become unstable, and analytic benchmarks are absent.

Looking ahead, we expect the benefits of the proposed framework to be even more substantial when the investment opportunity set is nonstationary—featuring time-varying or stochastic drifts and volatilities, factor-driven dynamics, or regime switches—a regime in which classical closed-form benchmarks are scarce and the structure of constrained optima is often unknown. In such settings, the local stagewise barrier solve remains lightweight and scalable, while value-based methods typically face severe challenges due to state-space expansion and action-space growth. 

Further extensions of the framework include smooth approximations of transaction costs, leverage or sector constraints, and path-dependent rules (e.g., consumption ratcheting) via barrier terms and state augmentation. The approach can also be generalized to incomplete-information environments and robust or distributionally robust formulations by modifying the Hamiltonian driver. On the algorithmic side, promising directions include adaptive barrier schedules, primal–dual projection schemes, and variance-reduced adjoint estimators.

Overall, projecting onto the Pontryagin/KKT manifold enables accurate, scalable, and constraint-respecting policies in high dimensions, remains effective even when classical solutions are unknown or fail to exist, and is poised to deliver even larger performance gains in realistically time-varying investment environments.

\section*{Acknowledgments}

This work was supported by the National Research Foundation of Korea (NRF) grant
funded by the Korea government (MSIT) (RS-2025-00562904 and RS-2024-00355646). 

\bibliographystyle{apalike}
\bibliography{gf_bib}

\newpage

\appendix

\section{Notation and Error Parameters Summary}\label{app:notation_summary}

Tables~\ref{tab:core_symbols}--\ref{tab:error_params} summarize the error/perturbation parameters
and the core symbols used throughout Sections~\ref{sec:PMP_constrained} and \ref{sec:ppgdpo}
and Appendices~\ref{app:proof_barrier_policy}--\ref{app:proof_ppgdpo}.

\begin{table}[H]
\centering
\caption{Core symbols, multipliers, and maps.}
\label{tab:core_symbols}
\begin{tabular}{@{}p{3.2cm}p{7.8cm}p{3.7cm}@{}}
\toprule
\textbf{Symbol} & \textbf{Meaning / Role} & \textbf{Where used} \\
\midrule
$\mathbf u := (\boldsymbol\pi, C)$ &
Control pair: risky-asset weights $\boldsymbol\pi$ and consumption $C$. &
Throughout; Secs.~\ref{sec:PMP_constrained}, \ref{sec:ppgdpo} \\
\addlinespace
$\mathcal H$ &
PMP Hamiltonian (unconstrained). &
Sec.~\ref{sec:PMP_unconstrained} \\
\addlinespace
$\widetilde{\mathcal H}_{\mathrm{bar}}$ &
Barrier-regularized Hamiltonian: $\mathcal H + \eta(1-\mathbf 1^\top\boldsymbol\pi) + \varepsilon_{\mathrm{bar}}\sum_j \ln\Gamma_j$. &
Sec.~\ref{sec:PMP_constrained}; App.~\ref{app:proof_barrier_policy} \\
\addlinespace
$\Gamma_j(\cdot)$ &
Inequality constraints; e.g., $\pi_i\ge 0$, $C_{\min}\le C \le C_{\max}$. &
Sec.~\ref{sec:PMP_constrained}; Assump.~\ref{ass:Gamma} \\
\addlinespace
$\eta$ &
Lagrange multiplier for the full-investment constraint $\mathbf 1^\top\boldsymbol\pi=1$. &
Secs.~\ref{sec:PMP_constrained}, \ref{sec:ppgdpo} \\
\addlinespace
$\nu = (\nu_j)_j$ &
Inequality multipliers; along the central path, $\nu_j = \varepsilon_{\mathrm{bar}}/\Gamma_j$. &
Sec.~\ref{sec:PMP_constrained}; App.~\ref{app:proof_barrier_policy} \\
\addlinespace
$\lambda$, $\mathbf Z$ &
Adjoints (costate and martingale term) of the PMP BSDE. &
Secs.~\ref{sec:PMP_unconstrained}, \ref{sec:ppgdpo}; App.~\ref{app:proof_bptt_pmp} \\
\addlinespace
$A := \partial_x \lambda$ &
Spatial derivative of $\lambda$ used to form $\mathbf Z$ at deployment. &
Sec.~\ref{sec:ppgdpo}; App.~\ref{app:proof_ppgdpo} \\
\addlinespace
$\widehat{\mathbf Z}(t,x,y;\boldsymbol\pi)$ &
Deployment-time relation $\; \widehat{\mathbf Z} = (\partial_x\widehat\lambda)\, x\,\widetilde{\boldsymbol v}^{\!\top}\boldsymbol\pi \;$ (``Markov identity''). &
Sec.~\ref{sec:ppgdpo} \\
\addlinespace
$H_{\mathrm{bar}}[\cdot]$ &
Reduced Hessian (tangent-space) operator of $\widetilde{\mathcal H}_{\mathrm{bar}}$; used in Newton–CG. &
Sec.~\ref{sec:PMP_constrained} \\
\addlinespace
$\mathbf u_{\mathrm{bar}}(\varepsilon_{\mathrm{bar}};\theta)$ &
Stagewise barrier maximizer given adjoints $\theta=(\lambda,A)$. &
App.~\ref{app:proof_ppgdpo} \\
\bottomrule
\end{tabular}
\end{table}

\begin{table}[H]
\centering
\caption{Error / perturbation parameters and residuals.}
\label{tab:error_params}
\begin{tabular}{@{}p{3.2cm}p{7.8cm}p{3.7cm}@{}}
\toprule
\textbf{Symbol} & \textbf{Meaning / Definition} & \textbf{Where used} \\
\midrule
$\varepsilon_{\mathrm{bar}}$ &
Log-barrier parameter for stagewise maximization; enforces interiority $\Gamma_j \ge c\,\varepsilon_{\mathrm{bar}}$ and adds diagonal curvature $\sim \varepsilon_{\mathrm{bar}}/\Gamma_j^2$; $\varepsilon_{\mathrm{bar}}\!\downarrow\!0$ $\Rightarrow$ KKT. &
Secs.~\ref{sec:PMP_constrained}, \ref{sec:ppgdpo}; Thms.~\ref{thm:barrier_policy}, \ref{thm:ppgdpo_gap_barrier}; Apps.~\ref{app:proof_barrier_policy}, \ref{app:proof_ppgdpo} \\
\addlinespace
$\varepsilon_{\mathrm{foc}}$ &
Warm-up Pontryagin FOC residual from activation-only training (measured in $L^{q,p}$). &
Sec.~\ref{sec:ppgdpo}; Thm.~\ref{thm:ppgdpo_gap_barrier}; App.~\ref{app:proof_ppgdpo} \\
\addlinespace
$\delta_{\mathrm{bptt}} \equiv \kappa_1\Delta t + \kappa_2/\sqrt{M}$ &
Adjoint estimation error from time discretization and Monte Carlo sampling. &
Sec.~\ref{sec:ppgdpo}; Thm.~\ref{thm:ppgdpo_gap_barrier}; App.~\ref{app:proof_ppgdpo} \\
\addlinespace
$\delta_{\mathrm{proj}}$ &
Stage~2 micro-solver residual (barrier KKT norm); choose $\delta_{\mathrm{proj}}\!\ll\! \varepsilon_{\mathrm{foc}}+\delta_{\mathrm{bptt}}$. &
Sec.~\ref{sec:ppgdpo}; Thm.~\ref{thm:ppgdpo_gap_barrier}; App.~\ref{app:proof_ppgdpo} \\
\addlinespace
$\varepsilon_k$ &
On-manifold stagewise FOC residual during training; summed as $\sum_k \varepsilon_k \Delta t$. &
App.~\ref{app:proof_bptt_pmp} \\
\bottomrule
\end{tabular}
\end{table}

\paragraph{Mixed time--probability norm.}
For a progressively measurable process $f:[0,T]\times(\Omega,\mathcal F,\mathbb P)\to\mathbb R^d$ on the filtered space $(\Omega,\mathcal F,(\mathcal F_t)_{t\in[0,T]},\mathbb P)$, we use the mixed norm
\[
\|f\|_{L^{q,p}}
:=\Bigg(\int_0^T \big(\,\mathbb E\!\left[\|f(t,\cdot)\|^p\right]\big)^{\!q/p}\,dt\Bigg)^{\!1/q}
=\|f\|_{L^{q}\!\left([0,T];\,L^{p}(\Omega;\mathbb R^d)\right)},
\]
with the usual modifications when $p=\infty$ or $q=\infty$ (essential suprema). Here $\|\cdot\|$ is the Euclidean norm on $\mathbb R^d$, and the product measure is $dt\otimes\mathbb P$. In particular, $\|\boldsymbol{\pi}^{\mathrm{proj}}-\boldsymbol{\pi}^*\|_{L^{q,p}}$ denotes this Bochner norm applied to the time‐indexed policy difference.

\section{Proof of Theorem~\texorpdfstring{\ref{thm:barrier_policy}}{}}\label{app:proof_barrier_policy}

We write $\varepsilon_{\mathrm{bar}}>0$ for the barrier parameter. Throughout, we adopt the sign convention
$\Gamma_j(\pi,C)\ge 0$ and the stagewise barrier Hamiltonian
\[
\mathcal H_{\mathrm{bar}}(\pi,C,\eta)
:= \mathcal H(\pi,C) + \eta\,(1-\mathbf 1^\top \pi) + \varepsilon_{\mathrm{bar}}\sum_{j=1}^m \ln \Gamma_j(\pi,C),
\]
where $\mathcal H$ is the PMP Hamiltonian at fixed $(t,x,\lambda,\mathbf Z)$ (suppressed in the notation). For later use,
let the (concave) Lagrangian of the constrained maximization be
\[
\mathcal L(\pi,C;\eta,\nu):=\mathcal H(\pi,C)+\eta(1-\mathbf 1^\top \pi)+\nu^\top \Gamma(\pi,C),
\qquad \Gamma:=(\Gamma_1,\ldots,\Gamma_m)^\top .
\]

\noindent\emph{About assumptions.}
In this appendix we rely only on Assumptions~\ref{ass:Gamma} and~\ref{ass:barrier_kkt}, in particular
Assumption~\ref{ass:barrier_kkt}(ii): LICQ, strict complementarity at $(\pi^*,C^*,\eta^*,\nu^*)$, and nonsingularity
of the bordered KKT Jacobian (strong metric regularity). 

\medskip\noindent
\textbf{Step 1 (Interior stationarity and central path).}
Any strictly interior stationary point $(\hat\pi_{\varepsilon_{\mathrm{bar}}},\hat C_{\varepsilon_{\mathrm{bar}}},\hat\eta_{\varepsilon_{\mathrm{bar}}})$ of $\mathcal H_{\mathrm{bar}}$ satisfies
\begin{align*}
\nabla_{\pi}\mathcal H(\hat\pi_{\varepsilon_{\mathrm{bar}}},\hat C_{\varepsilon_{\mathrm{bar}}})
- \hat\eta_{\varepsilon_{\mathrm{bar}}}\,\mathbf{1}
+ \sum_{j=1}^m \hat{\nu}_{j,\varepsilon_{\mathrm{bar}}}\,\nabla_{\pi}\Gamma_j(\hat\pi_{\varepsilon_{\mathrm{bar}}},\hat C_{\varepsilon_{\mathrm{bar}}}) &= 0, \\
\partial_C \mathcal H(\hat\pi_{\varepsilon_{\mathrm{bar}}},\hat C_{\varepsilon_{\mathrm{bar}}})
+ \sum_{j=1}^m \hat{\nu}_{j,\varepsilon_{\mathrm{bar}}}\,\partial_C \Gamma_j(\hat\pi_{\varepsilon_{\mathrm{bar}}},\hat C_{\varepsilon_{\mathrm{bar}}}) &= 0, \\
\mathbf{1}^\top \hat\pi_{\varepsilon_{\mathrm{bar}}} &= 1, \\
\hat{\nu}_{j,\varepsilon_{\mathrm{bar}}}\,\Gamma_j(\hat\pi_{\varepsilon_{\mathrm{bar}}},\hat C_{\varepsilon_{\mathrm{bar}}}) &= \varepsilon_{\mathrm{bar}} \quad (j=1,\dots,m),
\end{align*}
with strictly positive central–path multipliers $\hat{\nu}_{j,\varepsilon_{\mathrm{bar}}}:=\varepsilon_{\mathrm{bar}}/\Gamma_j(\hat\pi_{\varepsilon_{\mathrm{bar}}},\hat C_{\varepsilon_{\mathrm{bar}}})$.
By bounded superlevel sets of $\mathcal H_{\mathrm{bar}}$ (Assumption~\ref{ass:barrier_kkt}(i)), a maximizer exists for each fixed $\varepsilon_{\mathrm{bar}}>0$.
By LICQ and strong metric regularity in Assumption~\ref{ass:barrier_kkt}(ii), the interior stationary point is the
unique local maximizer near the KKT point.

\medskip\noindent
\textbf{Step 2 (IFT and $O(\varepsilon_{\mathrm{bar}})$ proximity).}
Set $u:=(\pi,C,\eta,\nu)$ and define the barrier KKT mapping
\[
G(u;\varepsilon_{\mathrm{bar}}):=
\begin{pmatrix}
\nabla_{\pi}\mathcal H(\pi,C)-\eta\,\mathbf 1 + J_{\pi}\Gamma(\pi,C)\,\nu\\[2pt]
\partial_C \mathcal H(\pi,C)+ \partial_C\Gamma(\pi,C)^\top\nu\\[2pt]
\mathbf 1^\top\pi-1\\[2pt]
\operatorname{diag}(\Gamma(\pi,C))\,\nu-\varepsilon_{\mathrm{bar}}\,\mathbf 1
\end{pmatrix}=0,
\]
where $J_{\pi}\Gamma:=[\nabla_{\pi}\Gamma_1\ \cdots\ \nabla_{\pi}\Gamma_m]$.
Let $u^\star=(\pi^*,C^*,\eta^*,\nu^*)$ be the KKT point at $\varepsilon_{\mathrm{bar}}=0$.
Assumption~\ref{ass:barrier_kkt}(ii) implies that $D_u G(u^\star;0)$ is nonsingular (on the equality tangent space),
hence by the Implicit Function Theorem there exists a unique $C^1$ \emph{central path} $u(\varepsilon_{\mathrm{bar}})$ for
$\varepsilon_{\mathrm{bar}}\in(0,\bar\varepsilon]$ with $G(u(\varepsilon_{\mathrm{bar}});\varepsilon_{\mathrm{bar}})=0$ and
\[
u(\varepsilon_{\mathrm{bar}})=u^\star + \Delta u_{\varepsilon_{\mathrm{bar}}}, \qquad
\|\Delta u_{\varepsilon_{\mathrm{bar}}}\| = O(\varepsilon_{\mathrm{bar}})\quad(\varepsilon_{\mathrm{bar}}\downarrow 0).
\]
Thus each component (in particular $\Delta\pi_{\varepsilon_{\mathrm{bar}}},\Delta C_{\varepsilon_{\mathrm{bar}}}$) is $O(\varepsilon_{\mathrm{bar}})$.

\medskip\noindent
\textbf{Step 3 (Componentwise $O(\varepsilon_{\mathrm{bar}})$ bounds).}
With the metric–regularity and local Lipschitz continuity in Assumptions~\ref{ass:barrier_kkt}(ii),(iii), we have
\[
\|\Delta\pi_{\varepsilon_{\mathrm{bar}}}\| + |\Delta C_{\varepsilon_{\mathrm{bar}}}| + |\hat\eta_{\varepsilon_{\mathrm{bar}}}-\eta^\ast|
+ \|\hat\nu_{\varepsilon_{\mathrm{bar}}}-\nu^\ast\| \;\le\; K\,\varepsilon_{\mathrm{bar}},
\]
for some $K>0$ independent of $\varepsilon_{\mathrm{bar}}$ on a neighborhood of $u^\ast$.

\medskip\noindent
\textbf{Step 4 (Lagrangian and Hamiltonian gaps).}
By local $L$-smoothness of $\nabla_{(\pi,C)} \mathcal L(\cdot;\eta^\ast,\nu^\ast)$ near $(\pi^\ast,C^\ast)$,
\[
0 \le \mathcal L(\pi^\ast,C^\ast;\eta^\ast,\nu^\ast) - \mathcal L(\hat\pi_{\varepsilon_{\mathrm{bar}}},\hat C_{\varepsilon_{\mathrm{bar}}};\eta^\ast,\nu^\ast)
\le \frac{L}{2}\big(\|\Delta\pi_{\varepsilon_{\mathrm{bar}}}\|^2 + |\Delta C_{\varepsilon_{\mathrm{bar}}}|^2\big)
= O(\varepsilon_{\mathrm{bar}}^2).
\]
For the Hamiltonian, use
\[
\mathcal H(\pi^\ast,C^\ast)-\mathcal H(\hat\pi_{\varepsilon_{\mathrm{bar}}},\hat C_{\varepsilon_{\mathrm{bar}}})
=
\Big[\mathcal L(\pi^\ast,C^\ast;\eta^\ast,\nu^\ast)-\mathcal L(\hat\pi_{\varepsilon_{\mathrm{bar}}},\hat C_{\varepsilon_{\mathrm{bar}}};\eta^\ast,\nu^\ast)\Big]
+ \nu^{\ast\top}\Gamma(\hat\pi_{\varepsilon_{\mathrm{bar}}},\hat C_{\varepsilon_{\mathrm{bar}}}),
\]
since $\mathbf{1}^\top \pi^\ast=\mathbf{1}^\top \hat\pi_{\varepsilon_{\mathrm{bar}}}=1$.
By strict complementarity along the central path,
$\Gamma_j(\hat\pi_{\varepsilon_{\mathrm{bar}}},\hat C_{\varepsilon_{\mathrm{bar}}})=\varepsilon_{\mathrm{bar}}/\hat\nu_{j,\varepsilon_{\mathrm{bar}}}$ for $j$ in the active set and $\hat\nu_{j,\varepsilon_{\mathrm{bar}}}\to\nu_j^\ast>0$,
so there exists $c>0$ with $\hat\nu_{j,\varepsilon_{\mathrm{bar}}}\ge c$ for small $\varepsilon_{\mathrm{bar}}$ and
\[
0 \le \nu^{\ast\top}\Gamma(\hat\pi_{\varepsilon_{\mathrm{bar}}},\hat C_{\varepsilon_{\mathrm{bar}}})
= \sum_{j\in\mathcal A}\nu_j^\ast \frac{\varepsilon_{\mathrm{bar}}}{\hat\nu_{j,\varepsilon_{\mathrm{bar}}}}
\le \Big(\tfrac{1}{c}\sum_{j\in\mathcal A}\nu_j^\ast\Big)\varepsilon_{\mathrm{bar}}
=: C_{H,1}\,\varepsilon_{\mathrm{bar}}.
\]
Hence
\[
0 \le \mathcal H(\pi^\ast,C^\ast)-\mathcal H(\hat\pi_{\varepsilon_{\mathrm{bar}}},\hat C_{\varepsilon_{\mathrm{bar}}})
\le C_{H,1}\,\varepsilon_{\mathrm{bar}} + C_{H,2}\,\varepsilon_{\mathrm{bar}}^2.
\]
Moreover, under the standard Lipschitz/linear–growth assumptions for the SDE coefficients
(covered by Assumption~\ref{ass:pmp_uncon}(ii)) and the local Lipschitz of $U'$
(Assumption~\ref{ass:bptt_pmp_reg}(ii)),
Grönwall’s inequality implies $\sup_{s\in[t,T]}\mathbb E|X^\ast_s-\hat X_{\varepsilon_{\mathrm{bar}},s}|=O(\varepsilon_{\mathrm{bar}})$, thus
\[
0 \le J(\pi^\ast,C^\ast) - J(\hat\pi_{\varepsilon_{\mathrm{bar}}},\hat C_{\varepsilon_{\mathrm{bar}}})
\le \kappa_1\,\varepsilon_{\mathrm{bar}}^2 + \kappa_2\,\varepsilon_{\mathrm{bar}}.
\]

\medskip\noindent
\textbf{Conclusion.} Steps 1–4 give the $O(\varepsilon_{\mathrm{bar}})$ policy accuracy, the $O(\varepsilon_{\mathrm{bar}}^2)$ Lagrangian gap,
and the mixed $O(\varepsilon_{\mathrm{bar}})+O(\varepsilon_{\mathrm{bar}}^2)$ Hamiltonian gap. Throughout, only
LICQ, strict complementarity, and nonsingularity of the bordered KKT Jacobian (Assumption~\ref{ass:barrier_kkt}(ii))
are used.

\begin{remark}[If strict complementarity is dropped]
If strict complementarity in Assumption~\ref{ass:barrier_kkt}(ii) is dropped,
the central path may approach $u^\ast$ only at rate $O(\sqrt{\varepsilon_{\mathrm{bar}}})$,
yielding $O(\sqrt{\varepsilon_{\mathrm{bar}}})$ policy error and an $O(\varepsilon_{\mathrm{bar}})$ instantaneous Hamiltonian gap.
\end{remark}

\section{Proof of Theorem~\texorpdfstring{\ref{thm:bptt_pmp_constrained}}{}}\label{app:proof_bptt_pmp}

We work in the activation-only setting: feasibility is enforced by smooth output maps
(softmax/simplex for $\boldsymbol{\pi}$, box/nonnegativity for $C$), and no log barrier
is used inside the rollout. Throughout we rely on Assumptions~\ref{ass:Gamma} and~\ref{ass:pmp_uncon};
the present appendix lists only the additional regularity specific to the BPTT--PMP argument.

\paragraph{Notation.}
We distinguish the explicit partial derivative $\partial_x$ (holding $(\boldsymbol{\pi},C)$ fixed)
from the total state derivative $\partial_X$ (which includes the dependence of $(\boldsymbol{\pi},C)$ on $X$
through the activation maps). Set
\[
R_k:=e^{-\rho t_k}U(C_k)\,\Delta t,\qquad
b_k:=X_k\,\boldsymbol{\pi}_k^\top \widetilde{\boldsymbol{\mu}}-C_k,\qquad
\sigma_k:=X_k\,\boldsymbol{\pi}_k^\top \widetilde{\boldsymbol{v}}.
\]
When $(\boldsymbol{\pi}_k,C_k)$ lies on the activation-induced feasible manifold
$\mathcal M_{t_k,X_k}$, we write the Riemannian (on-manifold) gradient as
\[
\nabla_{\!\mathcal M}\mathcal H
=\mathsf P_{T_{(\boldsymbol{\pi}_k,C_k)}\mathcal M}\,\nabla_{(\boldsymbol{\pi},C)}\mathcal H.
\]

\begin{assumption}[Regularity for BPTT--PMP (activation-only)]\label{ass:bptt_pmp_reg}
Work under Assumptions~\ref{ass:Gamma} and~\ref{ass:pmp_uncon}. In addition:
\begin{enumerate}[label=(\roman*), leftmargin=*]
\item \textbf{Driver Lipschitz and moment bounds.}
Assumption~\ref{ass:pmp_uncon}(ii) ensures the BSDE driver $-\partial_x\mathcal H$
is Lipschitz (and well-posed). In addition, the maps
\(
(x,\boldsymbol{\pi},C)\mapsto b(t,x;\boldsymbol{\pi},C),\;
(x,\boldsymbol{\pi})\mapsto \sigma(t,x;\boldsymbol{\pi})
\)
are locally Lipschitz on the (compact) training domain, uniformly in $t$.
Moreover, there exists $K_2>0$ such that
\(\sup_k \mathbb E[\,|X_k|^2+|\lambda_k|^2+\|\mathbf Z_k\|^2\,]\le K_2\).
\item \textbf{Utility and policy regularity.}
$U'$ is locally Lipschitz on the range of $C_\phi$. The policy outputs
$(\boldsymbol{\pi}_\theta,C_\phi)$ are $C^1$ and bounded, implemented via smooth activation maps
that send logits to $\mathcal M$ (simplex for $\boldsymbol{\pi}$; box/nonnegativity for $C$). On the training
domain the Jacobians are uniformly bounded: $\exists\,\bar K>0$ with
$\|\partial_X(\boldsymbol{\pi}_k,C_k)\|\le \bar K$ for all $k$.
\item \textbf{Stagewise on-manifold $\varepsilon_k$-stationarity.}
At step $k$,
\[
\big\|\nabla_{\!\mathcal M}\,\mathcal H(t_k,X_k;\boldsymbol{\pi}_k,C_k,\lambda_{k+1},\mathbf Z_{k+1})\big\|
\ \le\ \varepsilon_k,\qquad
\varepsilon_{\mathrm{tot}}:=\sum_k \varepsilon_k\,\Delta t<\infty.
\]
\item \textbf{Control interpolation.}
Let $\mathbf u^{\Delta t}(t)$ be the piecewise-constant interpolation $\mathbf u_k$ on $[t_k,t_{k+1})$.
As $\Delta t\to0$, $\mathbf u^{\Delta t}\to \mathbf u$ in $L^2([0,T]\times\Omega)$ for some progressively
measurable $\mathbf u=(\boldsymbol{\pi},C)$ induced by the same networks; in particular,
$X^{\Delta t}\Rightarrow X$ and the stochastic integrals are uniformly integrable.
\end{enumerate}
\end{assumption}

\begin{proof}[Proof of Theorem~\ref{thm:bptt_pmp_constrained}]
We proceed in four steps.

\medskip\noindent
\textbf{Step 1 (Exponential--Euler remainder).}
The exponential--Euler update admits an Euler--Maruyama form
\begin{equation}\label{eq:EM_form_app_B}
X_{k+1}
= X_k + b_k\,\Delta t + \sigma_k\cdot \Delta \mathbf{W}_k + r^{X}_k,
\end{equation}
with one–step remainders satisfying, for a constant $C>0$ independent of $k,\Delta t$,
\[
\mathbb E\!\left[|r^{X}_k|\,\big|\,\mathcal F_{t_k}\right]\le C\,\Delta t^{3/2},
\qquad
\mathbb E\!\left[|r^{X}_k|^2\,\big|\,\mathcal F_{t_k}\right]\le C\,\Delta t^{2}.
\]
This follows from the first–order It\^o–Taylor expansion and the Lipschitz/linear–growth bounds in
Assumption~\ref{ass:bptt_pmp_reg}(i)–(ii).

\medskip\noindent
\textbf{Step 2 (On–manifold envelope identity with explicit algebra).}
Define
\[
Q_k\ :=\ \frac{1}{\Delta t}\frac{\partial R_k}{\partial X_k}
\;+\;\lambda_{k+1}\,\partial_X b_k\;+\;(\partial_X\sigma_k)\,\mathbf Z_{k+1}.
\]
We claim
\begin{equation}\label{eq:envelope_identity}
Q_k\;=\;\partial_x \mathcal H\!\left(t_k,X_k;\boldsymbol{\pi}_k,C_k,\lambda_{k+1},\mathbf Z_{k+1}\right)
\;+\;\big\langle\nabla_{(\boldsymbol{\pi},C)}\mathcal H,\ \partial_X(\boldsymbol{\pi}_k,C_k)\big\rangle.
\end{equation}
Indeed, using
\(
\partial_C\mathcal H=e^{-\rho t_k}U'(C_k)-\lambda_{k+1},\;
\partial_{\boldsymbol{\pi}}\mathcal H
=X_k\lambda_{k+1}\widetilde{\boldsymbol{\mu}}
+X_k\widetilde{\boldsymbol v}\,\mathbf Z_{k+1},
\)
and the identities
\[
\partial_X b_k-\partial_x b_k
=\partial_{(\boldsymbol{\pi},C)}b_k\,\partial_X(\boldsymbol{\pi}_k,C_k)
= \big(X_k\widetilde{\boldsymbol{\mu}},\ -1\big)\,\partial_X(\boldsymbol{\pi}_k,C_k),
\]
\[
\partial_X \sigma_k-\partial_x \sigma_k
=\partial_{\boldsymbol{\pi}}\sigma_k\,\partial_X\boldsymbol{\pi}_k
= X_k\,\widetilde{\boldsymbol v}\,\partial_X\boldsymbol{\pi}_k,
\]
together with
\(
\frac{1}{\Delta t}\frac{\partial R_k}{\partial X_k}
=e^{-\rho t_k}U'(C_k)\,\partial_X C_k,
\)
a direct collection of terms yields exactly \eqref{eq:envelope_identity}.
By Cauchy–Schwarz on the tangent space and Assumption~\ref{ass:bptt_pmp_reg}(ii)–(iii),
\[
\big|\langle\nabla_{(\boldsymbol{\pi},C)}\mathcal H,\ \partial_X(\boldsymbol{\pi}_k,C_k)\rangle\big|
\ \le\ \|\nabla_{\!\mathcal M}\mathcal H\|\;\|\partial_X(\boldsymbol{\pi}_k,C_k)\|
\ \le\ \bar K\,\varepsilon_k,
\]
whence
\begin{equation}\label{eq:Q_is_driver_plus_eps}
Q_k\;=\;\partial_x \mathcal H(t_k,X_k;\boldsymbol{\pi}_k,C_k,\lambda_{k+1},\mathbf Z_{k+1})
\;+\;O(\varepsilon_k),
\end{equation}
with constants depending only on local Lipschitz bounds and $\bar K$.

\medskip\noindent
\textbf{Step 3 (Discrete adjoint recursion with controlled remainder).}
Define the discrete adjoints
\[
\lambda_k := \frac{\partial J_{\mathrm e}}{\partial X_k},
\qquad
\mathbf Z_{k+1} := \frac{1}{\Delta t}\,\mathbb E\!\left[\lambda_{k+1}\,\Delta \mathbf W_k \,\big|\, \mathcal F_{t_k}\right].
\]
By the $L^2$ projection property there exists an \emph{orthogonal remainder} $\xi_{k+1}$ with
\[
\mathbb E[\xi_{k+1}\mid\mathcal F_{t_k}]=0,\qquad
\mathbb E[\xi_{k+1}\,\Delta\mathbf W_k^\top\mid\mathcal F_{t_k}]=\boldsymbol 0,
\]
such that
\[
\lambda_{k+1}
=\mathbb E[\lambda_{k+1}\mid\mathcal F_{t_k}] + \mathbf Z_{k+1}^\top \Delta\mathbf W_k + \xi_{k+1}.
\]
By the BPTT chain rule,
\[
\lambda_k \;=\; \frac{\partial R_k}{\partial X_k}
\;+\; \mathbb{E}\!\left[\,\lambda_{k+1}\,\frac{\partial X_{k+1}}{\partial X_k}\,\Big|\,\mathcal{F}_{t_k}\right].
\]
From \eqref{eq:EM_form_app_B},
\(
\frac{\partial X_{k+1}}{\partial X_k}
= 1 + \partial_X b_k\,\Delta t + \partial_X \sigma_k \cdot \Delta \mathbf W_k
  + \tilde r^X_k,
\)
with $\mathbb E[|\tilde r^X_k|\mid\mathcal F_{t_k}]=O(\Delta t^{3/2})$
and $\mathbb E[|\tilde r^X_k|^2\mid\mathcal F_{t_k}]=O(\Delta t^{2})$.
Using the above martingale decomposition of $\lambda_{k+1}$ and
$\mathbb E[\Delta\mathbf W_k\Delta\mathbf W_k^\top\mid\mathcal F_{t_k}]=\Delta t\,I$,
we obtain
\[
\mathbb E[\lambda_k-\lambda_{k+1}\mid\mathcal F_{t_k}]
=
\Delta t\,\bigg(\frac{1}{\Delta t}\frac{\partial R_k}{\partial X_k}+\lambda_{k+1}\partial_X b_k
+(\partial_X\sigma_k)\mathbf Z_{k+1}\bigg)
+O(\Delta t^{3/2}),
\]
where the terms involving $\xi_{k+1}$ drop out in conditional expectation and the remaining contribution is absorbed into the remainder below. Invoking \eqref{eq:Q_is_driver_plus_eps} yields
\[
\lambda_k - \lambda_{k+1}
=
\partial_x \mathcal H\!\left(t_k, X_k; \boldsymbol{\pi}_k, C_k, \lambda_{k+1}, \mathbf Z_{k+1}\right)\,\Delta t
\;-\; \mathbf Z_{k+1}^\top \Delta \mathbf W_k 
\;+\; r_k ,
\]
with
\[
\mathbb E[r_k\mid\mathcal F_{t_k}]=O(\Delta t^{3/2}+\varepsilon_k\Delta t),\qquad
\mathbb E[|r_k|^2\mid\mathcal F_{t_k}]=O(\Delta t^{2}+\varepsilon_k^2\Delta t^{2}),
\]
after absorbing the contributions of $\tilde r_k^X$ and $\xi_{k+1}$ into $r_k$.

\medskip\noindent
\textbf{Step 4 (Convergence to the PMP adjoint).}
Let $(\lambda^{\Delta t},\mathbf Z^{\Delta t})$ be the piecewise-constant interpolants of
$(\lambda_k,\mathbf Z_k)$. For any bounded test process $\varphi$,
\[
\mathbb E\bigg[\sum_k \Big(\lambda_k-\lambda_{k+1}
+\mathbf Z_{k+1}^\top \Delta\mathbf W_k
+\partial_x\mathcal H(t_k,X_k;\mathbf u_k,\lambda_{k+1},\mathbf Z_{k+1})\,\Delta t\Big)\,\varphi_k\bigg]
= O\!\big(\Delta t^{1/2}+\varepsilon_{\mathrm{tot}}\big).
\]
By Assumption~\ref{ass:bptt_pmp_reg}(i) the driver $-\partial_x\mathcal H$ is Lipschitz in $(\lambda,\mathbf Z)$,
and the moment bound yields uniform $L^2$ stability (see, e.g., \cite{ma1999forward,yong2012stochastic}).
Together with the control interpolation in Assumption~\ref{ass:bptt_pmp_reg}(iv),
standard Euler-scheme arguments for Lipschitz BSDEs imply that any weak $L^2$ limit point
$(\lambda,\mathbf Z)$ of $(\lambda^{\Delta t},\mathbf Z^{\Delta t})$ solves
\[
d\lambda_t
=
-\partial_x \mathcal H\!\left(t, X_t; \boldsymbol{\pi}_t, C_t, \lambda_t, \mathbf Z_t\right)\,dt
+ \mathbf Z_t^\top d\mathbf W_t .
\]
Uniqueness of the BSDE solution then gives full $L^2$ convergence
$(\lambda^{\Delta t},\mathbf Z^{\Delta t})\to(\lambda,\mathbf Z)$ as $\Delta t\to0$ and $\varepsilon_{\mathrm{tot}}\to0$.
\emph{Equivalently, the discrete recursion above is the explicit Euler discretization of the adjoint BSDE with a vanishing local truncation error, whence $L^2$ convergence follows.}
\end{proof}

\begin{remark}[No barrier in the BSDE driver]
Because feasibility is enforced directly by the activation mappings during training, no barrier terms
appear in the BSDE driver. If desired, a log-barrier projection can be applied at deployment
(Appendix~\ref{app:proof_ppgdpo}); it does not change the BPTT--PMP correspondence established here.
\end{remark}

\section{Proof of Theorem~\texorpdfstring{\ref{thm:ppgdpo_gap_barrier}}{}}\label{app:proof_ppgdpo}

\paragraph{Notation and unification.}
We write \(\varepsilon_{\mathrm{foc}}\) for the warm-up Pontryagin first-order residual (activation-only training),
\(\varepsilon_{\mathrm{bar}}>0\) for the log-barrier parameter used in Stage~2 projection,
and \(\delta_{\mathrm{proj}}\) for the numerical residual of the Stage~2 micro-solver (measured in a fixed
Euclidean KKT residual norm; cf.\ Remark~\ref{rem:uniformity_norm}).
Set \(\theta:=(\lambda,A)\) with \(A:=\partial_x\lambda\), and denote the true adjoints by
\(\theta^*:=(\lambda^*,A^*)\).
At projection time we use the Markov relation
\[
\mathbf Z(t,x,y;\boldsymbol\pi)\ :=\ A(t,x,y)\, x\,\widetilde{\boldsymbol v}^{\!\top}\boldsymbol\pi.
\]
We write \(\mathbf u:=(\boldsymbol\pi,C)\) and \(\mathbf u^*:=(\boldsymbol\pi^*,C^*)\) for the control pair and the
PMP/KKT-optimal pair, respectively.

\medskip

\begin{assumption}[Barrier projection setup (uniform version)]\label{ass:barrier_setup_appendix}
Fix a compact deployment domain in \((t,x,y)\). This is a uniform (domain-wide) strengthening of Assumption~\ref{ass:barrier_kkt}. For all sufficiently small
\(\varepsilon_{\mathrm{bar}}\in(0,\bar\varepsilon]\) and all \(\theta\) near \(\theta^*\),
the stagewise barrier problem
\[
\mathbf u_{\mathrm{bar}}(\varepsilon_{\mathrm{bar}};\theta)
\ :=\
\arg\max_{\substack{\Gamma(t,x,y;\mathbf u)>\boldsymbol 0\\ \mathbf 1^\top\boldsymbol\pi=1}}
\!\!\Big\{\mathcal H\big(t,x,y;\mathbf u,\lambda,\mathbf Z(t,x,y;\boldsymbol\pi)\big)
+\eta(1-\mathbf 1^\top\boldsymbol\pi)
+\varepsilon_{\mathrm{bar}}\sum_{j} \ln\Gamma_j(t,x,y;\mathbf u)\Big\}
\]
admits a unique \emph{strict} local maximizer whose reduced Hessian of the \emph{barrier objective}
\(\widetilde{\mathcal H}_{\mathrm{bar}}\) on the equality tangent space is uniformly positive definite, 
and the interiority is uniform:
\(\Gamma_j(t,x,y;\mathbf u_{\mathrm{bar}})\ge c\,\varepsilon_{\mathrm{bar}}\) for some \(c\in(0,1)\) and all \(j\).
All constants may be chosen \emph{uniformly} on the deployment domain and for
\(\varepsilon_{\mathrm{bar}}\in(0,\bar\varepsilon]\).
\end{assumption}

Assumption~\ref{ass:barrier_setup_appendix} uniformizes the stagewise regularity from the main text
(see also Assumptions~\ref{ass:Gamma} and~\ref{ass:barrier_kkt}). It implies strong metric regularity of the
barrier KKT system and a uniform interior central path.

\begin{proposition}[Adjoint stability under activation-only feasibility]\label{prop:adjoint_stability}
Let \(\varepsilon_{\mathrm{foc}}\) be the warm-up FOC residual (activation-only PG-DPO), and define
\(\delta_{\mathrm{bptt}}:=\kappa_1\Delta t+\kappa_2/\sqrt{M}\) with \(\Delta t\) the rollout step and \(M\) Monte Carlo paths.
Then there exist constants \(a_1,a_2>0\) such that
\[
\|\widehat\theta-\theta^*\|_{L^{q,p}}
\ \le\
a_1\,\varepsilon_{\mathrm{foc}}\;+\;a_2\,\delta_{\mathrm{bptt}}.
\]
The constants are uniform on the deployment domain.
\end{proposition}

\paragraph{Barrier KKT map and metric regularity.}
Let \(\zeta:=(\boldsymbol\pi,C,\eta,\nu)\) with \(\nu\in\mathbb R^m\) and \(\mathbf u:=(\boldsymbol\pi,C)\).
Define the barrier KKT residual
\[
G(\zeta;\theta,\varepsilon_{\mathrm{bar}})=
\begin{pmatrix}
\nabla_{\boldsymbol\pi}\mathcal H(t,x,y;\mathbf u,\lambda,\mathbf Z)-\eta\,\mathbf 1 + J_{\boldsymbol\pi}\Gamma(t,x,y;\mathbf u)\,\nu\\
\partial_C\mathcal H(t,x,y;\mathbf u,\lambda,\mathbf Z)+\partial_C\Gamma(t,x,y;\mathbf u)^{\!\top}\nu\\
\mathbf 1^\top\boldsymbol\pi-1\\
\mathrm{diag}\big(\Gamma(t,x,y;\mathbf u)\big)\,\nu-\varepsilon_{\mathrm{bar}}\mathbf 1
\end{pmatrix}.
\]
By Assumption~\ref{ass:barrier_setup_appendix}, the Jacobian \(D_{\!\zeta}G\) is uniformly invertible on the equality
tangent space along the interior path, hence the system is \emph{strongly metrically regular}. In particular,
there exists a \(C^1\) central path \(\zeta_{\mathrm{bar}}(\varepsilon_{\mathrm{bar}};\theta)\) with a \emph{uniform} inverse bound.

\begin{proof}[Proof of Theorem~\ref{thm:ppgdpo_gap_barrier}]
Fix \((t,x,y)\) and suppress it in notation. Let \(\mathbf u^{\mathrm{proj}}\) denote the Stage~2 output at
\((\widehat\theta,\varepsilon_{\mathrm{bar}})\), and let \(\mathbf u_{\mathrm{bar}}(\varepsilon_{\mathrm{bar}};\theta)\) be the barrier maximizer at \((\theta,\varepsilon_{\mathrm{bar}})\).

\medskip\noindent
\textbf{Step 1 (Triangle decomposition).}
\begin{equation}\label{eq:tri_decomp_ppgdpo}
\|\mathbf u^{\mathrm{proj}}-\mathbf u^*\|
\ \le\
\underbrace{\|\mathbf u^{\mathrm{proj}}-\mathbf u_{\mathrm{bar}}(\varepsilon_{\mathrm{bar}};\widehat\theta)\|}_{\mathrm{(I)}}
+\underbrace{\|\mathbf u_{\mathrm{bar}}(\varepsilon_{\mathrm{bar}};\widehat\theta)-\mathbf u_{\mathrm{bar}}(\varepsilon_{\mathrm{bar}};\theta^*)\|}_{\mathrm{(II)}}
+\underbrace{\|\mathbf u_{\mathrm{bar}}(\varepsilon_{\mathrm{bar}};\theta^*)-\mathbf u^*\|}_{\mathrm{(III)}}.
\end{equation}

\medskip\noindent
\textbf{Step 2 (Micro-solver residual \(\delta_{\mathrm{proj}}\)).}
Assume the Stage~2 iterate \(\hat\zeta\) satisfies
\(\|G(\hat\zeta;\widehat\theta,\varepsilon_{\mathrm{bar}})\|_{\mathsf{KKT}}\le \delta_{\mathrm{proj}}\),
where \(\|\cdot\|_{\mathsf{KKT}}\) is any fixed norm equivalent to the Euclidean norm on the stacked
(primal stationarity, equality feasibility, perturbed complementarity) residual.
Strong metric regularity of \(G\) yields
\[
\|\hat\zeta-\zeta_{\mathrm{bar}}(\varepsilon_{\mathrm{bar}};\widehat\theta)\|\ \le\ K_{\mathrm{res}}\ \delta_{\mathrm{proj}},
\qquad\Longrightarrow\qquad
\mathrm{(I)}\ \le\ K_{\mathrm{res}}\ \delta_{\mathrm{proj}}.
\]

\medskip\noindent
\textbf{Step 3 (Parameter sensitivity in \(\theta\)).}
Apply the Implicit Function Theorem to \(G(\zeta;\theta,\varepsilon_{\mathrm{bar}})=0\) (equivalently to the reduced
system in \(\mathbf u\) after eliminating \(\eta\) by a \(1\times1\) Schur complement). Uniform invertibility on the
tangent space implies
\[
\big\|\zeta_{\mathrm{bar}}(\varepsilon_{\mathrm{bar}};\widehat\theta)-\zeta_{\mathrm{bar}}(\varepsilon_{\mathrm{bar}};\theta^*)\big\|
\ \le\ K_{\mathrm{sens}}\,\|\widehat\theta-\theta^*\|,
\qquad\Rightarrow\qquad
\mathrm{(II)}\ \le\ K_{\mathrm{sens}}\,\|\widehat\theta-\theta^*\|.
\]
Combining with Proposition~\ref{prop:adjoint_stability} yields
\[
\mathrm{(II)}\ \le\ K_{\mathrm{sens}}\,(a_1\varepsilon_{\mathrm{foc}}+a_2\delta_{\mathrm{bptt}}).
\]

\medskip\noindent
\textbf{Step 4 (Barrier central-path bias).}
By Theorem~\ref{thm:barrier_policy} (barrier vs.\ KKT), uniformly on the deployment domain,
\[
\mathrm{(III)}\ \le\ K_{\mathrm{bar}}\ \varepsilon_{\mathrm{bar}}.
\]

\medskip\noindent
\textbf{Step 5 (Collecting bounds; vector \(\to\) portfolio block).}
Substituting the bounds for \(\mathrm{(I)}\)–\(\mathrm{(III)}\) into \eqref{eq:tri_decomp_ppgdpo} and taking \(L^{q,p}\)-norms,
\[
\|\mathbf u^{\mathrm{proj}}-\mathbf u^*\|_{L^{q,p}}
\ \le\
C_{\mathrm{proj}}\Big(\varepsilon_{\mathrm{foc}}+\delta_{\mathrm{bptt}}+\delta_{\mathrm{proj}}+\varepsilon_{\mathrm{bar}}\Big),
\]
for a uniform constant \(C_{\mathrm{proj}}>0\) depending only on model bounds and the tangent-space
strong concavity modulus (hence independent of \(\varepsilon_{\mathrm{foc}},\Delta t,M,\delta_{\mathrm{proj}},\varepsilon_{\mathrm{bar}}\)).
Since \(\|\boldsymbol\pi^{\mathrm{proj}}-\boldsymbol\pi^*\|\le \|\mathbf u^{\mathrm{proj}}-\mathbf u^*\|\), the policy-gap bound for \(\boldsymbol\pi\) follows.

\medskip\noindent
\textbf{Step 6 (Hamiltonian gap; sharpened bound).}
Set $\Delta\mathbf u:=\mathbf u^{\mathrm{proj}}-\mathbf u_{\mathrm{bar}}(\varepsilon_{\mathrm{bar}};\theta^*)$.
Recall the exact identity (equality constraint holds at both points)
\[
\mathcal H(\mathbf u^*)-\mathcal H(\mathbf u^{\mathrm{proj}})
=\big[\mathcal L(\mathbf u^*;\eta^*,\nu^*)-\mathcal L(\mathbf u^{\mathrm{proj}};\eta^*,\nu^*)\big]
\;+\;{\nu^*}^{\!\top}\Gamma(\mathbf u^{\mathrm{proj}}),
\]
where $\mathcal L(\cdot;\eta^*,\nu^*)=\mathcal H(\cdot)+\eta^*(1-\mathbf 1^\top\boldsymbol\pi)+{\nu^*}^{\!\top}\Gamma(\cdot)$.
By local $L$–smoothness of $\nabla_{\mathbf u}\mathcal L(\cdot;\eta^*,\nu^*)$ near the KKT point,
the Lagrangian drop is quadratically bounded:
\[
0\ \le\ \mathcal L(\mathbf u^*;\eta^*,\nu^*)-\mathcal L(\mathbf u^{\mathrm{proj}};\eta^*,\nu^*)
\ \le\ C_{\mathrm{tan}}\;\|\Delta\mathbf u\|^2 .
\]
For the complementarity term, decompose
\[
{\nu^*}^{\!\top}\Gamma(\mathbf u^{\mathrm{proj}})
\ \le\ {\nu^*}^{\!\top}\Gamma\!\big(\mathbf u_{\mathrm{bar}}(\varepsilon_{\mathrm{bar}};\theta^*)\big)
\;+\;\big\| \nu^*\big\|\,\mathrm{Lip}(\Gamma)\,\|\Delta\mathbf u\|.
\]
Along the central path at $(\theta^*,\varepsilon_{\mathrm{bar}})$, strict complementarity implies
$\nu_{\varepsilon,j}\to \nu^*_j>0$ for $j\in\mathcal A$ and $\nu^*_j=0$ for $j\notin\mathcal A$. Since
$\nu_{\varepsilon,j}\,\Gamma_j(\mathbf u_{\mathrm{bar}})=\varepsilon_{\mathrm{bar}}$ componentwise, we obtain
\[
{\nu^*}^{\!\top}\Gamma\!\big(\mathbf u_{\mathrm{bar}}(\varepsilon_{\mathrm{bar}};\theta^*)\big)
=\sum_{j\in\mathcal A}\nu^*_j\,\frac{\varepsilon_{\mathrm{bar}}}{\nu_{\varepsilon,j}}
\ \le\ C_{H,1}\,\varepsilon_{\mathrm{bar}}
\]
for all sufficiently small $\varepsilon_{\mathrm{bar}}$, where $C_{H,1}$ depends only on lower bounds of $\nu_{\varepsilon,j}$ on $\mathcal A$.
Combining the bounds and using Steps~2--3 (triangle decomposition), we get
\[
0\ \le\ \mathcal H(\mathbf u^*)-\mathcal H(\mathbf u^{\mathrm{proj}})
\ \le\ C_{H,1}\,\varepsilon_{\mathrm{bar}}
\;+\;C_{\mathrm{tan}}\|\Delta\mathbf u\|^2
\;+\;\|\nu^*\|\mathrm{Lip}(\Gamma)\,\|\Delta\mathbf u\|.
\]
By Steps~2--3,
$\|\Delta\mathbf u\|\le K_{\mathrm{res}}\,\delta_{\mathrm{proj}}
+K_{\mathrm{sens}}\big(a_1\varepsilon_{\mathrm{foc}}+a_2\delta_{\mathrm{bptt}}\big)$.
Hence, for some $C_{H,2},C_{H,3}>0$ (uniform on the deployment domain),
\[
\mathcal H(\mathbf u^*)-\mathcal H(\mathbf u^{\mathrm{proj}})
\ \le\ C_{H,1}\,\varepsilon_{\mathrm{bar}}
\;+\;C_{H,2}\big(\varepsilon_{\mathrm{foc}}+\delta_{\mathrm{bptt}}+\delta_{\mathrm{proj}}\big)
\;+\;C_{H,3}\big(\varepsilon_{\mathrm{foc}}^{\,2}+\delta_{\mathrm{bptt}}^{\,2}+\delta_{\mathrm{proj}}^{\,2}\big)\ .
\]
In particular, applying Young’s inequality to the linear term yields the coarser but often convenient quadratic form
\[
\mathcal H(\mathbf u^*)-\mathcal H(\mathbf u^{\mathrm{proj}})
\ \le\ \widetilde C_{H,1}\,\varepsilon_{\mathrm{bar}}
\;+\;\widetilde C_{H,2}\big(\varepsilon_{\mathrm{foc}}^{\,2}+\delta_{\mathrm{bptt}}^{\,2}+\delta_{\mathrm{proj}}^{\,2}\big),
\]
and if one prefers a purely quadratic bound (as in the main-text summary), it further implies
$\ \mathcal H(\mathbf u^*)-\mathcal H(\mathbf u^{\mathrm{proj}})
\le \widehat C_{H}\big(\varepsilon_{\mathrm{bar}}^{2}+\varepsilon_{\mathrm{foc}}^{\,2}+\delta_{\mathrm{bptt}}^{\,2}+\delta_{\mathrm{proj}}^{\,2}\big)$
whenever $\varepsilon_{\mathrm{bar}}$ is not asymptotically dominant.

\medskip\noindent
Combining Steps~1–6 proves the stated policy-gap bound. The Hamiltonian gap admits the sharper mixed-order bound boxed above; the quadratic summary bound in the main text follows by Young’s inequality.
\end{proof}

\begin{remark}[Uniformity of constants and residual norm]\label{rem:uniformity_norm}
The constants \(K_{\mathrm{res}},K_{\mathrm{sens}},K_{\mathrm{bar}},C_{\mathrm{proj}}\) can be chosen uniformly on the
deployment domain and for all sufficiently small \(\varepsilon_{\mathrm{bar}}\in(0,\bar\varepsilon]\), provided the tangent-space
strong concavity modulus and the interiority margin \(\Gamma_j\ge c\,\varepsilon_{\mathrm{bar}}\) are uniform
(Assumption~\ref{ass:barrier_setup_appendix}). The projection residual \(\delta_{\mathrm{proj}}\) is measured in any fixed
norm equivalent to the Euclidean norm on the stacked barrier KKT residual; changing the norm only
rescales \(K_{\mathrm{res}}\) by a constant factor.
\end{remark}

\begin{remark}[On dimension dependence]
While the constants depend on model coefficients and the tangent-space strong concavity modulus,
the proof uses no grid structure and the deployment-time solve is of size \(n{+}2\); hence the per-step
cost scales linearly in the number of risky assets \(n\). If the modulus deteriorates with \(n\), the
constants reflect that dependence; our uniformity assumption prevents explosion of constants on
the relevant deployment set.
\end{remark}

\end{document}